%% file: Recursive_Marginal_Quantization.tex
\documentclass[11pt,a4paper,notitlepage]{article}
\usepackage{amsmath}
\usepackage{amsthm}
\usepackage{amssymb}
\usepackage{mathrsfs}
\usepackage{lipsum}
\usepackage[square]{natbib}
\usepackage{graphicx}
\usepackage{hyperref}
\hypersetup{
    colorlinks=true,  
    linkcolor=blue,   
    citecolor=blue
}
\usepackage{authblk}
\usepackage{bbm}
\usepackage{color}
\usepackage{pgf}
\usepackage{a4wide}	

\newcommand{\R}{\mathbb{R}}                         
\newcommand{\Np}{\mathbb{N^+}}                      

\newcommand{\E}[1]{\mathbb{E}\left[#1\right]}       
\newcommand{\Ebig}[1]{\mathbb{E}\Big[#1\Big]}       
\renewcommand{\P}{\mathbb{P}}                       
\newcommand{\sigalg}[1]{\mathscr{#1}}               
\newcommand{\F}{\sigalg{F}}                         
\newcommand{\triple}{(\Omega,\F,\P)}                
\newcommand{\ind}[1]{\mathbb{I}_{\left\{#1\right\}}}

\newcommand{\tr}{t\in[0,T]}                         
\newcommand{\filt}{(\F_t)_{\tr}}                    
\newcommand{\basis}{(\Omega,\F,\filt,\P)}           

\newcommand{\dx}[2]{#1'(#2)}
\newcommand{\ddx}[2]{#1''(#2)}

\newcommand{\cw}[1]{\gamma^{#1}}        
\newcommand{\cwt}[2]{\gamma_{#1}^{#2}}  
\renewcommand{\rm}[1]{r^{#1-}}          
\newcommand{\rp}[1]{r^{#1+}}            
\newcommand{\Rm}[2]{r_{#1}^{#2-}}
\newcommand{\Rp}[2]{r_{#1}^{#2+}}

\newcommand{\affm}[2]{m_{#1}^{#2}}
\newcommand{\affc}[2]{c_{#1}^{#2}}

\newcommand{\Xq}{\widehat{X}}           
\newcommand{\Xd}{\widetilde{X}}         
\newcommand{\Delt}{\Delta t}            

\newcommand{\U}{\mathcal{U}}

\newcommand{\mP}{\mathbf{P}}
\newcommand{\mM}{\mathbf{M}}
\newcommand{\mf}{\mathbf{f}}
\newcommand{\mm}{\mathbf{m}}
\newcommand{\mc}{\mathbf{c}}
\newcommand{\mJ}{\mathbf{j}}
\renewcommand{\mp}{\mathbf{p}}
\renewcommand{\mho}{\mathbf{h}_{\mathsf{off}}}
\newcommand{\mhm}{\mathbf{h}_{\mathsf{main}}}
\newcommand{\hprod}{\circ}
\newcommand{\mGamma}{\mathbf{\Gamma}}

\newcommand{\sgn}{\,\mathrm{sgn}}

\theoremstyle{plain}

\title{Recursive Marginal Quantization of Higher-Order Schemes}
\author[1,2]{Thomas A. McWalter\thanks{Correspondence: tom@analytical.co.za}}
\author[1]{Ralph Rudd}
\author[1,3]{J\"{o}rg Kienitz}
\author[1,4]{Eckhard Platen}
\affil[1]{Department of Actuarial Science and the African Collaboration for Quantitative Finance and Risk Research, University of Cape Town}
\affil[2]{Department of Finance and Investment Management, University of Johannesburg}
\affil[3]{Fachbereich Mathematik und Naturwissenschaften,
Bergische Universit\"{a}t Wuppertal}
\affil[4]{Finance Discipline Group and School of Mathematical and Physical Sciences, University of Technology Sydney}

\begin{document}

\maketitle

\begin{abstract}
Quantization techniques have been applied in many challenging finance applications, including pricing claims with path dependence and early exercise features, stochastic optimal control, filtering problems and efficient calibration of large derivative books. Recursive Marginal Quantization of the Euler scheme has recently been proposed as an efficient numerical method for evaluating functionals of solutions of stochastic differential equations. This method involves recursively quantizing the conditional marginals of the discrete-time Euler approximation of the underlying process. By generalizing this approach, we show that it is possible to perform recursive marginal quantization for two higher-order schemes: the Milstein scheme and a simplified weak order 2.0 scheme. As part of this generalization a simple matrix formulation is presented, allowing efficient implementation. We further extend the applicability of recursive marginal quantization by showing how absorption and reflection at the zero boundary may be incorporated, when this is necessary. To illustrate the improved accuracy of the higher order schemes, various computations are performed using geometric Brownian motion and its generalization, the constant elasticity of variance model. For both processes, we show numerical evidence of improved weak order convergence and we compare the marginal distributions implied by the three schemes to the known analytical distributions. By pricing European, Bermudan and Barrier options, further evidence of improved accuracy of the higher order schemes is demonstrated.
\end{abstract}

\section{Introduction}
\label{Sec: Introduction}
\input{Sections/Introduction.tex}

\section{Vector Quantization}
\label{Sec: Vector Quantization}
\input{Sections/VQ}

\section{Recursive Marginal Quantization}
\label{Sec: RMQ}
\input{Sections/RMQ.tex}

\section{Higher-order RMQ Schemes}
\label{Sec: Higher Order}
\input{Sections/HigherOrder.tex}

\section{The Zero Boundary}
\label{Sec: Absorption and Reflection}
\input{Sections/AbsorptionReflection.tex}

\section{Pricing}
\label{Sec: Pricing}
\input{Sections/Pricing.tex}

\section{Conclusion}
\label{Sec: Conclusion}
\input{Sections/Conclusion.tex}

\clearpage

\bibliographystyle{abbrvnat}
\bibliography{RMQ_References}

\end{document}

%% file: Sections/Introduction.tex
Quantization techniques have been shown to be effective in a number of quantitative finance applications. In particular, this includes the pricing of contingent claims with path dependency and early exercise \citep{PagesWilbertz2012, sagna2010pricing, Bormettietal2016}, stochastic control problems \citep{PagesPhamPrintems2004} and nonlinear filtering \citep{PagesPham2005}. To improve numerical efficiency, \cite{pages2015recursive} introduced a technique known as recursive marginal quantization. This technique has been shown to be effective for fast calibration of large derivative books by \cite{callegaro2014pricing,Callegaroetal2015a}, another challenging application in finance. This paper aims to significantly improve the accuracy of recursive marginal quantization.

Quantization is a lossy compression technique that produces a discrete representation of a signal using less information than the original. The technique originated in the field of signal compression, but has found application in fields as far-reaching  as signal processing, pattern recognition, data mining, integration theory and, more recently, numerical probability. For general overviews of the mathematics and applications of quantization see \cite{du1999centroidal} and \cite{pagesintroduction}.

Vector quantization of probability distributions was formalized in the work of \cite{graf2000foundations} and has been applied to the field of mathematical finance since its inception. It is a technique for optimally representing a continuous distribution by a discrete distribution, where a measure of the `distance' between the two, called the distortion, is minimized. The distortion is most commonly measured using the squared Euclidean error.

The application of vector quantization to the solution of finance-related problems generally proceeds by discretising time and then quantizing the corresponding marginal distributions of the system of stochastic differential equations specific to the problem. The quantized grids and their associated weights are then used to compute the expectations required in pricing contingent claims (including claims with early exercise) or for performing the optimizations required in stochastic control problems \citep{pages2004optimal}. Due to the reliance on Lloyd's Algorithm \citep{lloyd1982least}, or variants thereof, these approaches generally incur a heavy computational burden.

A more efficient approach for the single-factor case has recently been proposed by \cite{pages2015recursive}. Known as recursive marginal quantization (RMQ), it makes use of a Newton-Raphson iteration to quantize the Euler-Maruyama \citep{Maruyama1955} updates of the underlying SDE. This technique has been used to provide fast calibration of a local volatility model by \cite{callegaro2014pricing,Callegaroetal2015a}, and extended for use with two factor SDEs and applied to stochastic volatility models \citep{callegaro2015pricing}.

In the present work, the RMQ algorithm is generalized, allowing the implementation of schemes of higher order than the Euler-Maruyama scheme. We now provide an overview of the rest of the paper.

Section Two provides a review of vector quantization (VQ) as applied to probability distributions. In particular, we strive to simultaneously provide a precise, concise and intuitive description of the methodology. The resulting algorithm is presented using a matrix formulation, allowing for efficient implementation. The section concludes by showing examples of vector quantization applied to the Gaussian and noncentral chi-squared distributions.

The third section introduces recursive marginal quantization applied to stochastic differential equations. Our formulation of the problem is presented in more generality than the original formulation by \cite{pages2015recursive}.
A matrix formulation, which demonstrates the connection with Markov chains, is provided, allowing easy and efficient implementation.

In the section that follows we extend the RMQ algorithm to higher-order updates, specifically the \cite{mil1975approximate} scheme and a simplified weak order 2.0  Taylor scheme of \cite{kloeden1999numerical}. Geometric Brownian motion (GBM) and the constant elasticity of variance (CEV) process serve as examples to illustrate the improved error in the quantized marginal distributions and the improved weak order convergence.

When performing a Monte Carlo simulation of a discrete-time approximation of a process, non-negativity of the solution is usually enforced by implementing absorption or reflection. Under certain circumstances this is also required when using RMQ. In Section Five we present the modifications of the RMQ algorithm necessary to ensure an absorbing or reflecting boundary at zero. These modifications allow the RMQ algorithm to be applied to the CEV process for parameter sets that would otherwise be problematic under the original formulation.

Section Six presents numerical results of the application of the three RMQ schemes to option pricing. European, barrier and Bermudan options are priced under the GBM and CEV models. Where possible, the results are compared to available closed-form solutions, otherwise they are compared to high-resolution finite difference or Monte Carlo implementations.

One of the goals of this paper has been to produce a more general and easily accessible introduction to the theory of VQ and RMQ. The primary contribution is, however, the more general formulation of RMQ that enables the systematic extension of the work of \cite{pages2015recursive}. The paper concludes with a discussion of ongoing work.

%% file: Sections/VQ.tex
\begin{figure}[b!]
    \begin{center}
        \includegraphics[width=0.6\columnwidth]{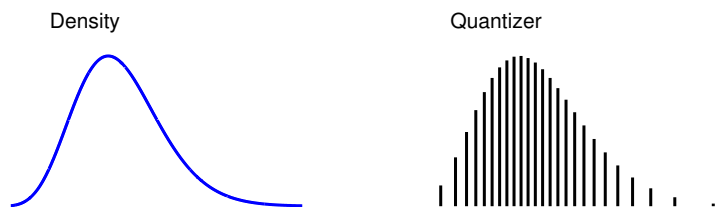}
    \end{center}
    \caption{The continuous probability density function on the left is quantized on the right, with probabilities represented on the vertical axis.}\label{Fig: Quantized Density}
\end{figure}

Vector quantization is a lossy compression technique that provides a way to encode a vector space using a discrete subspace. While the technique is applicable more generally, we shall only consider the quantization of one dimensional distributions. The vector quantization problem we aim to address in this section may be specified intuitively as follows:
\begin{quote}
    Find the discrete distribution that ``best'' represents the continuous distribution function associated with a random variable $X$.
\end{quote}
This is depicted in the Figure \ref{Fig: Quantized Density}, which shows a density function of a continuous random variable and its corresponding quantized version. Here, for ease of visualization, we have chosen to plot the probability density function (of the continuous random variable) and the probability mass function of the quantizer instead of the continuous and discrete distribution functions.

We now provide a more rigorous specification of the problem. Let $X$ be a continuous random variable, taking values in $\R$, and defined on a probability space $\triple$. The above question may be reposed as:
\begin{quote}
    How does one optimally approximate $X$, in a least-squares sense, by a discrete random variable $\widehat{X}:\Omega\rightarrow\Gamma$, where $\Gamma$ is a finite set of elements in $\R$?
\end{quote}
The reason that quantization is useful is that it allows efficient approximations of expectations of functionals $H(X)$ of $X$, that is,
\[
    \E{H(X)}=\int_{\R}H(x)\,d\P(X\leq x)\approx\sum_{\gamma\in\Gamma}H(\gamma)\P\big(\widehat{X}=\gamma\big). 
\]
Here, for example, $H$ may be the discounted payoff of a financial claim and $\P$ the risk-neutral probability measure.

Consider the approximation of $X$ given by $\hat{X}$, a discrete random vector defined as the nearest-neighbour projection of $X$ onto $\Gamma=\{\cw{1},\cw{2},\ldots,\cw{N}\}$, a set of distinct points in $\R$ with finite cardinality $N\in\Np$. We shall refer to $\Gamma$ as the \emph{quantizer} and its elements as \emph{codewords}. The \emph{nearest neighbor projection operator} $\pi_\Gamma:\R\rightarrow\Gamma$ is defined as
\begin{align*}
    \pi_\Gamma(X)=\big\{\cw{i}\in\Gamma\,\big|\,\,\|X-\cw{i}\|\leq \|X-\cw{j}\|\text{ for all $j=1,\ldots,N;$ }&\text{where equality}\\
    &\text{holds only for $i<j$}\big\}.
\end{align*}
Here, $\|\cdot\|$ is the Euclidean norm. Associated with the quantizer, the \emph{regions} $R_i(\Gamma)\subset\R$ are the subset of values of $X$ that are mapped to each codeword $\cw{i}$:
\[
    R_i(\Gamma)=\big\{x\in\R\,\big|\,\,\pi_\Gamma(x)=\cw{i}\big\}.
\]
These are also known as \emph{Voronoi regions}. For the sake of brevity we shall use $R_i$ to refer to $R_i(\Gamma)$ when it is clear that the quantizer we are referring to is $\Gamma$. The set of regions $\{R_i\}_{i=1}^N$ is called a \emph{tessellation} of $\R$, and has the following properties:
\[
    R_i\cap R_j=\emptyset\text{ for }i\neq j\qquad\text{and}\qquad\cup_{i=1}^NR_i=\R.
\]
Since we are working in one dimension, the regions $R_i$ may be defined directly as $R_i=\{x\,|\,\rm{i}<x\leq\rp{i}\}$ with
\[
    \rm{i}=\frac{\cw{i-1}+\cw{i}}2\qquad\text{and}\qquad \rp{i}=\frac{\cw{i}+\cw{i+1}}2,
\]
for $1\leq i\leq N$, where, by definition, $\rm{1}=-\infty$ and $\rp{N}=\infty$. If the distribution under consideration is not defined over the whole real line, then $\rm{1}$ and $\rp{N}$ are adjusted to reflect the interval of support. Figure \ref{Regions} shows a simple graphical representation of these regions.

\input{Figures/regions.TpX}

With these definitions in place, we are now in a position to precisely define the optimization problem. We wish to find the quantizer $\Gamma$ such that $\widehat{X}=\pi_\Gamma(X)$ best approximates $X$. The sense in which the quantizer $\Gamma$ is ``best'' is specified by the \emph{distortion function}
\begin{align}
    D(\Gamma)&=\E{\|X-\widehat{X}\|^2}\notag\\
    &=\int_{\R}\|x-\pi_\Gamma(x)\|^2\,d\P(X\leq x)\notag\\
    &=\sum_{i=1}^N\int_{R_i(\Gamma)}\|x-\cw{i}\|^2\,d\P(X\leq x).\label{Eq: VQ Distortion}
\end{align}
We require the $\Gamma$ that minimizes $D(\Gamma)$. The probability weights then follow directly as a result of the nearest neighbor projection operator, i.e., $\P(\widehat{X}=\cw{i})=\P(X\in R_i)$. A necessary condition on the optimal $\Gamma$ is that the gradient of the distortion function is zero, that is, $\nabla D(\Gamma)=\bar{0}$, where the elements of $\nabla D(\Gamma)$ are given by
\[
    \frac{\partial D(\Gamma)}{\partial \cw{i}}=2\left(\int_{R_i(\Gamma)}(\cw{i}-x)\,d\P(X\leq x)\right),
\]
for $1\leq i\leq N$.
Intuitively, this means that the first moment conditioned on the outcomes of a region equals the respective codeword. Thus, one way to solve this system of equations is to set up a fixed-point iteration using the above expression for the gradient. This is the basis for Lloyd's algorithm which starts with an initial guess for the quantizer, $\Gamma^{(0)}$, and generates successive updates, $\Gamma^{(n+1)}$, with the new codewords, $\cwt{n+1}{i}$, computed as the centroids of the regions associated with the previous updates $\Gamma^{(n)}$:
\[
    \cwt{n+1}{i}=\frac{\int_{R_i(\Gamma^{(n)})}x\,d\P(X\leq x)}{\int_{R_i(\Gamma^{(n)})}d\P(X\leq x)},
\]
for $1\leq i\leq N$ and $0\leq n<n_{\mathrm{max}}$.

Another approach is to represent the quantizer by the column vector $\mGamma$, derive the Hessian, $\nabla^2D(\mGamma)$, and compute the updated estimates of the quantizer using an iterative Newton-Raphson method
\[
    \mGamma^{(n+1)}=\mGamma^{(n)}-\left[\nabla^2D\left(\mGamma^{(n)}\right)\right]^{-1}\nabla D\left(\mGamma^{(n)}\right)
\]
for $0\leq n<n_{\mathrm{max}}$. We now develop this approach further.

Suppose $f_X$ and $F_X$ are the PDF and CDF of $X$, respectively. We define the $p$-th lower partial expectation as
\[
    M_X^p(x)=\E{X^p\ind{X<x}},
\]
where $M_X^0(X)=F_X(x)$ represents the distribution function of $X$. Then, direct integration of the distortion function \eqref{Eq: VQ Distortion} gives
\begin{align*}
    D(\Gamma)&=\sum_{i=1}^N\int_{\rm{i}}^{\rp{i}} \|x-\cw{i}\|^2f_X(x)\,d x\\
    &=\sum_{i=1}^N\Big[M_X^2(\rp{i})-M_X^2(\rm{i})
    -2\cw{i}\left(M_X^1(\rp{i})-M_X^1(\rm{i})\right)\\
    &\qquad\quad\quad+(\cw{i})^2\left(F_X(\rp{i})-F_X(\rm{i})\right)\Big].
\end{align*}
Consequently, the elements of the vector $\nabla D(\mGamma)$ are given by
\[
    \frac{\partial D(\Gamma)}{\partial \cw{i}}=2\cw{i}\left(F_X(\rp{i})-F_X(\rm{i})\right)-2\left( M_X^1(\rp{i})-M_X^1(\rm{i})\right),
\]
for $1\leq i\leq N$.

Similarly, the tridiagonal Hessian matrix, $\nabla^2D(\mGamma)$, may be computed. It has diagonal elements given by
\begin{align*}
    \frac{\partial^2D(\Gamma)}{\partial (\cw{i})^2}&=2\left(F_X(\rp{i})-F_X(\rm{i})\right)
    +\tfrac12\left(f_X(\rp{i})(\cw{i}-\cw{i+1})
    +f_X(\rm{i})(\cw{i-1}-\cw{i})\right),\\
\intertext{and super- and sub-diagonal elements given by}
    \frac{\partial^2D(\Gamma)}{\partial\cw{i}\partial \cw{i+1}}&=\tfrac12f_X(\rp{i})(\cw{i}-\cw{i+1})\\
\intertext{and}
    \frac{\partial^2D(\Gamma)}{\partial\cw{i}\partial
    \cw{i-1}}&=\tfrac12f_X(\rm{i})(\cw{i-1}-\cw{i}),
\end{align*}
respectively. Note that the quantities required to compute a Newton-Raphson iteration (i.e., the gradient and Hessian) only require the PDF, CDF and first lower partial expectation to be known. The second lower partial expectation is required only if one wishes to compute a numerical estimate of the distortion.

\subsection{Efficient Implementation}
\label{Sec: VQ Efficient Implementation}

We now provide a matrix formulation of the above Newton iteration intended to aid efficient implementation. As stated previously, the quantizer is represented by a column vector $\mGamma$. This vector and three other column vectors required are defined by
\begin{align*}
    [\mGamma]_i&=\gamma^i, & [\mM]_i&=M_X^1(\rp{i})-M_X^1(\rm{i}), & & 1\leq i\leq N,\\
    [\mf]_i&=f_X(\rp{i}), & [\Delta\mGamma]_i&=\cw{i+1}-\cw{i}, & & 1\leq i\leq N-1.
\intertext{Note that the last two vectors are one element shorter than the first two. The row vector of probabilities $\mp$ is defined as } 
     & & [\mp]_i&=\P(\Xq=\cw{i})\\
     & & &=\P(X\in R_i(\Gamma))\\
     & & &=F_X(\rp{i})-F_X(\rm{i}),& & 1\leq i\leq N.
\end{align*}
Defining $\mp$ as a row vector is convenient since the expectation of a functional $H$ applied to the quantizer is
\begin{equation}
    \E{H(X)}=\sum_{i=1}^NH(\cw{i})\P\big(\widehat{X}=\cw{i}\big)=\mp H(\mGamma), \label{Eq: Expectation of a Functional}
\end{equation}
where $H$ is applied element-wise to $\mGamma$. Moreover, this will be compatible with a Markov chain formulation of the recursive marginal quantization technique presented later.

Using these vectors the gradient of the distortion function is then
\[
    \nabla D(\mGamma)=2\mGamma\circ\mp^\top-2\mM,
\]
where $\circ$ indicates the element-wise Hadamard product.

The super- and sub-diagonal (or off-diagonal) entries of the Hessian matrix $\nabla^2D(\mGamma)$ are given by the length-$(N-1)$ row vector
\[
    \mho=-\frac12[\mf\circ\Delta\mGamma]^\top,
\]
with the main diagonal given by
\[
    \mhm=2\mp+[\mho|0]+[0|\mho],
\]
where the copies of the $\mho$ vector are appended and prepended with a zero. It is now straightforward to set up the Newton-Raphson iteration in terms of the quantities.

\subsection{Examples}
In this section we apply the above methodology to the Gaussian distribution and the noncentral chi-squared distribution with one degree of freedom. The latter is important for the higher-order recursive marginal quantization schemes that we explore later in the paper.

\subsubsection{The Standard Normal Distribution}
\begin{figure}
    \begin{center}
        \includegraphics[width=0.7\columnwidth]{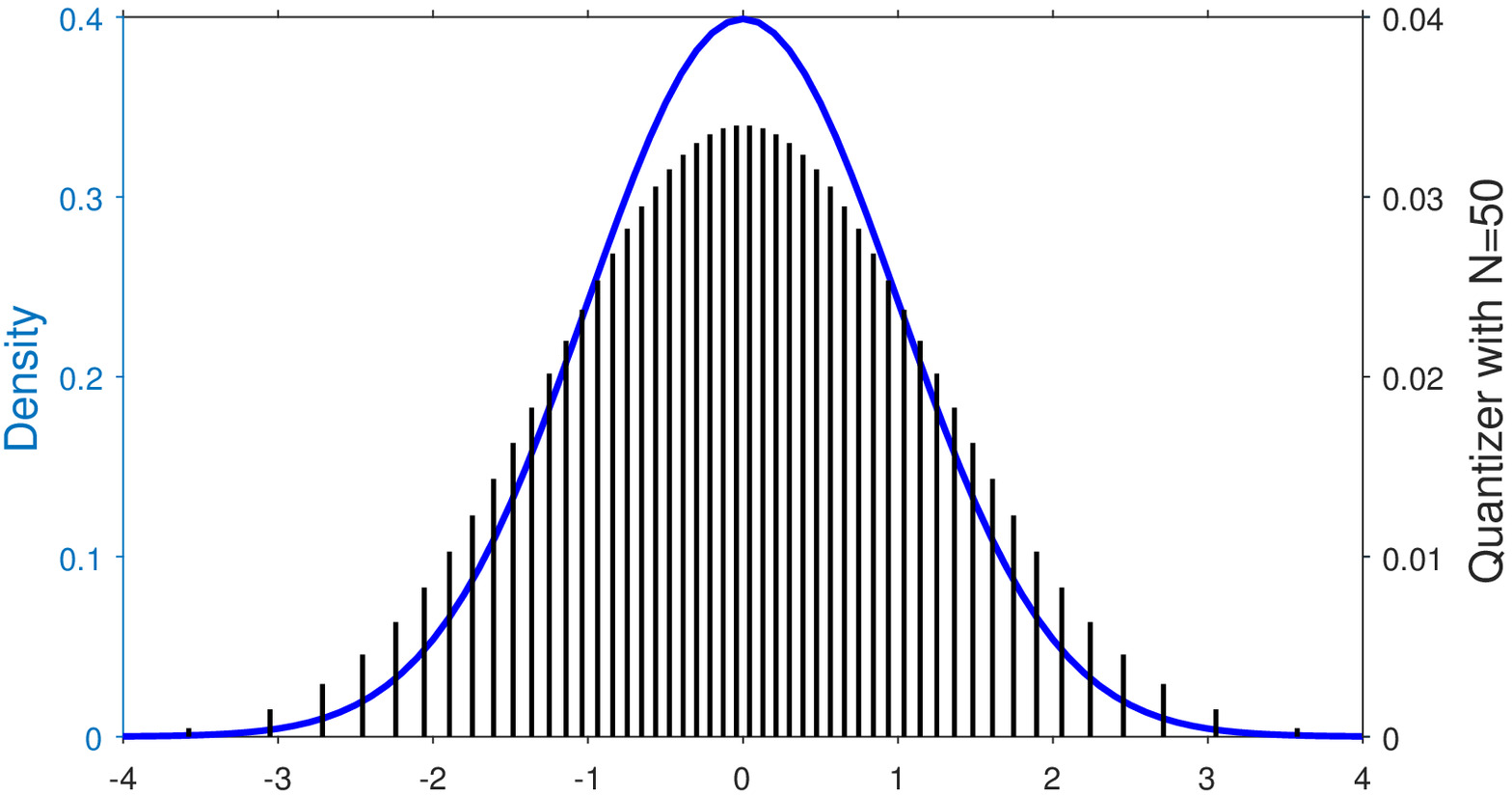}
    \end{center}
    \caption{Vector Quantization of the Standard Normal Distribution}\label{Fig: VQ Normal}
\end{figure}

When $X$ is a standard normal random variable we have
\begin{align*}
    f_X(x)&=\phi(x)\\
    F_X(x)&=\Phi(x)\\
    M_X^1(x)&=-\frac1{\sqrt{2\pi}}e^{-\frac{x^2}2}=-\phi(x),
\end{align*}
where $\phi(\cdot)$ and $\Phi(\cdot)$ are the standard normal PDF and CDF, respectively. Here, a good guess for the initial quantizer $\mGamma^{(0)}$ is
\[
    \cw{n}=\frac{5.5n}{N+1}-2.75,
\]
for $1\leq n\leq N$. Figure \ref{Fig: VQ Normal} shows the quantizer of cardinality $N=50$, using this initial guess, after $n_{\mathrm{max}}=20$ Newton-Raphson iterations. At first glance, it is tempting to think of the quantizer (represented by bars) as one might think of a histogram and suspect that it does not adequately capture the features of the density because it has the ``incorrect shape''. This is, however, a misleading analogy since a histogram represents the probability of realising a random variable in equally sized intervals on the $x$-axis. The quantizer, however, accumulates the probability mass over the Voronoi regions associated with each of the codewords and, as these regions are larger in the tails than in the body of the distribution, proportionally more probability mass is accumulated for codewords in the tails than for codewords in the body.

\subsubsection{The Noncentral Chi-squared Distribution}

While, in general, the noncentral chi-squared distribution must be specified using Bessel functions, this is not the case when the degree of freedom equals one. In particular, consider the random variable $X=(Z+\mu)^2$, where $Z\sim\mathscr{N}(0,1)$. Then $X\sim\chi^{\prime2}(1,\lambda)$, is a noncentral chi-squared distributed with one degree of freedom and noncentrality parameter $\lambda=\mu^2$. Moreover, on $x\in\R^+$, we have
\begin{align*}
    f_X(x)&=\frac1{2\sqrt{x}}\left(\phi(x^+)+\phi(x^-)\right)\\
    F_X(x)&=\Phi(x^+)-\Phi(x^-)\\[1.7mm]
    M_X^1(x)&=(1+\lambda)\left(\Phi(x^+)-\Phi(x^-)\right)+\phi(x^+)x^--\phi(x^-)x^+,
\intertext{where $\phi(\cdot)$ and $\Phi(\cdot)$ are the standard normal PDF and CDF, respectively, and}
    x^{\pm}&=\pm\sqrt{x}-\sqrt{\lambda}.
\end{align*}
This means that we may express the noncentral chi-squared distribution with one degree of freedom using the standard normal PDF and CDF, thus allowing efficient computation of a quantization scheme. This will be important for computational efficiency when we implement higher-order RMQ schemes later in the paper.

When implementing VQ, care must be taken when evaluating these functions at left and right limits. To ensure convergence we set $f_X(0)=F_X(0)=M_X^1(0)=0$, $f_X(\infty)=0$, $F_X(\infty)=1$ and $M_X^1(\infty)=1+\lambda$. Of course, all three functions are zero when $x$ is negative. A good initial guess for $\mGamma^{(0)}$ is given by
\[
    \cw{n}=\begin{cases}
    \left(\frac{(3+\sqrt{\lambda})n}{N}\right)^2&\text{for $\sqrt{\lambda}<2.5$}\\
    \left(\frac{5n}{N+1}-2.5+\sqrt{\lambda}\right)^2&\text{for $\sqrt{\lambda}\geq2.5$,}
    \end{cases}
\]
for $1\leq n\leq N$.

\begin{figure}
    \begin{center}
        \includegraphics[width=0.7\columnwidth]{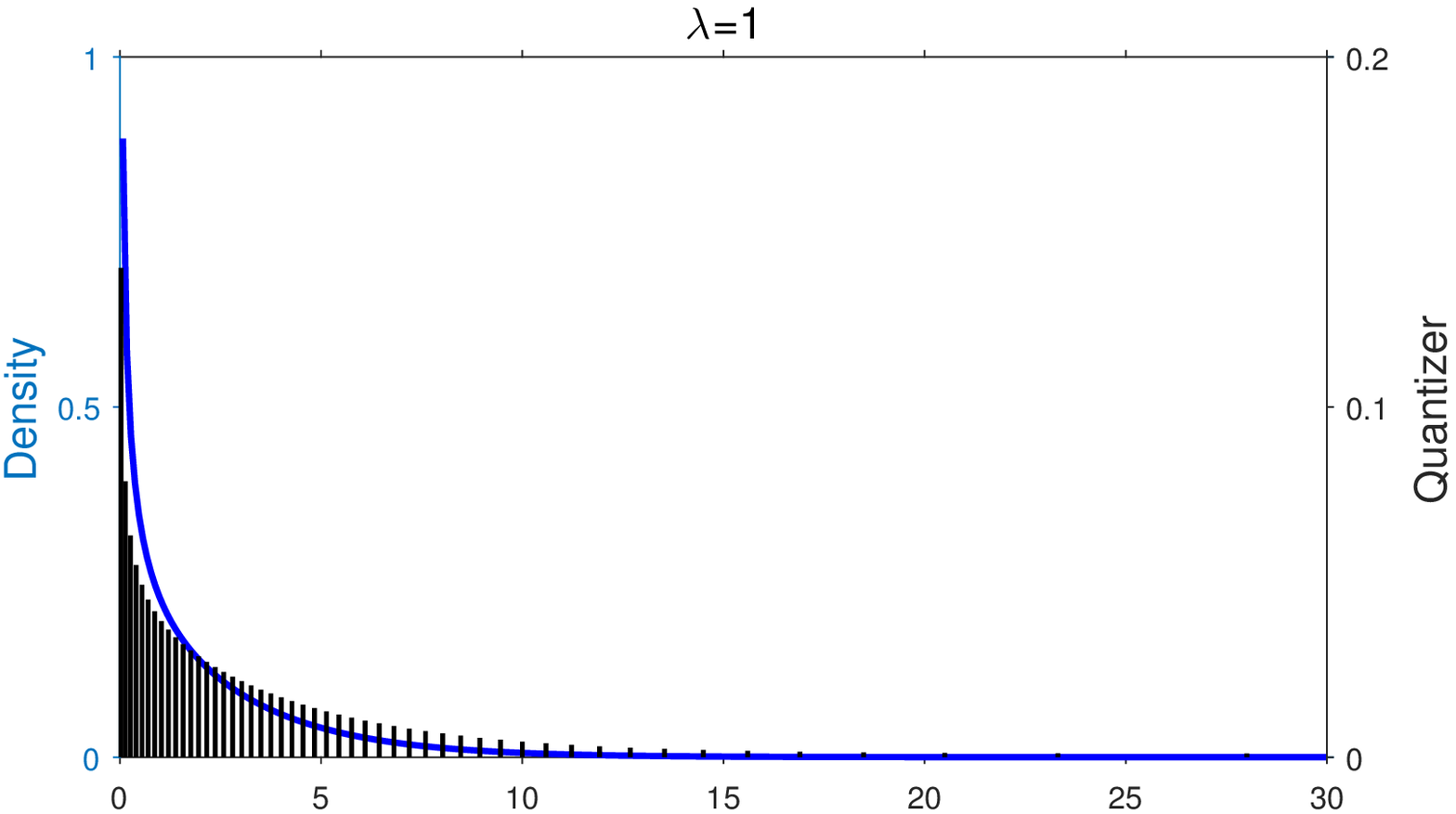}
    \end{center}
    \begin{center}
        \includegraphics[width=0.7\columnwidth]{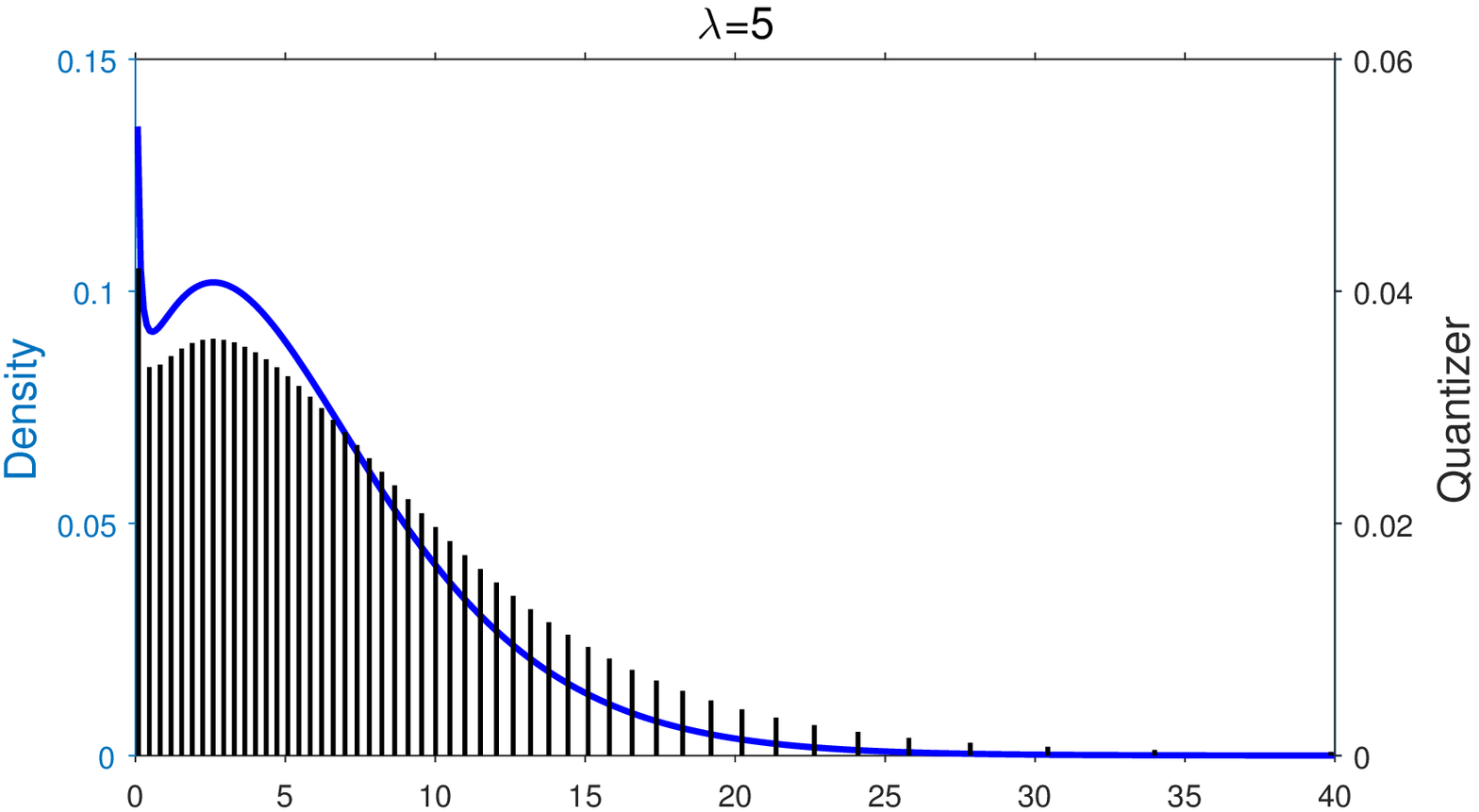}
    \end{center}
    \begin{center}
        \includegraphics[width=0.7\columnwidth]{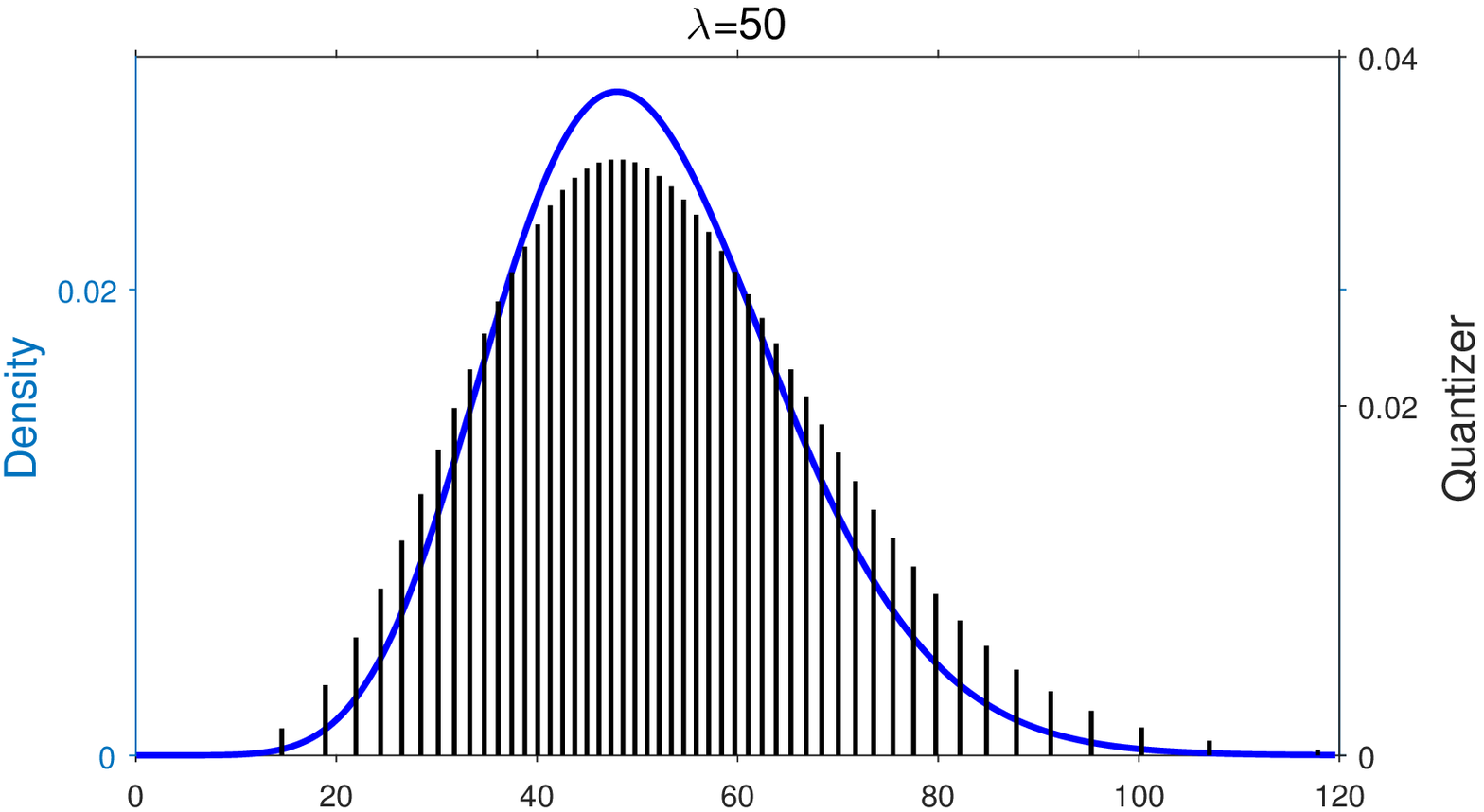}
    \end{center}
    \caption{Three examples of the noncentral chi-squared distribution with one degree of freedom for different noncentrality parameters.}\label{Fig: VQ Chi-Squared}
\end{figure}

Figure \ref{Fig: VQ Chi-Squared} shows three examples of quantizers of cardinality $N=50$ for the noncentral chi-squared distribution with one degree of freedom for a range of noncentrality parameters. Note that for certain values of the noncentrality parameter (e.g.\ the central panel) the distribution is bimodal while for larger values it resembles a Gaussian distribution.

%% file: Sections/RMQ.tex
Consider the continuous-time diffusion specified by the stochastic differential equation
\begin{equation}
    dX_t=a(X_t)\,dt+b(X_t)\,dW_t,\qquad X_0=x_0\in\R,\label{Eq: SDE}
\end{equation}
defined on the filtered probability space $\basis$ with $a$ and $b$ sufficiently smooth and regular functions to ensure the existence of a weak solution. The question of interest is:
\begin{quote}
    How does one optimally approximate $X_{t_k}:\Omega\rightarrow\R$, for some time discretisation point $t_k\in[0,T]$, when the distribution of $X_{t_k}$ is unknown?
\end{quote}

Usually this is achieved by performing a Monte Carlo experiment using a discrete-time approximation scheme for the SDE, the simplest scheme being the Euler-Maruyama \citep{Maruyama1955} update
\begin{align*}
    \Xd_{k+1}&=\Xd_{k}+a(\Xd_k)\Delt +b(\Xd_k)\sqrt{\Delt}Z_{k+1}\\
    &=:\U(\Xd_{k},Z_{k+1}),
\end{align*}
for $0\leq k<K$, where $\Delt=T/K$ and independent $Z_{k+1}\sim\mathscr{N}(0,1)$, with initial value $\Xd_0=x_0$. The innovation of \cite{pages2015recursive} was to show that a recursive procedure based on quantizing these updates is possible.

Since $\Xd_{1}$ has a Gaussian distribution, it is possible to use vector quantization to obtain $\Gamma_1$, an optimal quantization grid for the first step of the above scheme. One must, however, find a way to quantize the successive (marginal) distributions of $\Xd_{k+1}$. Given knowledge of the distribution of $\Xd_{k}$, the distortion of the quantizer $\Gamma_{k+1}$ may be written as
\begin{align}
    \widetilde{D}\big(\Gamma_{k+1}\big)&=\E{\big\|\Xd_{k+1}-\Xq_{k+1}\big\|^2}\notag\\
    &=\E{\E{\left.\big\|\Xd_{k+1}-\Xq_{k+1}\big\|^2\,\right|\,\Xd_k}}\notag\\
    &=\E{\E{\left.\big\|\U(\Xd_k,Z_{k+1})-\Xq_{k+1}\big\|^2\,\right|\,\Xd_k}}\notag\\
    &=\int_{\R}\E{\big\|\U(x,Z_{k+1})-\Xq_{k+1}\big\|^2}\!d\P(\Xd_k\leq x).\label{Eq: Integrated Distortion}
\end{align}
Unfortunately, we do not know the exact distribution of $\Xd_{k}$ for $k>1$.
The main result of \cite{pages2015recursive} shows that if one uses the previously quantized distribution of $\Xq_{k}$, instead of the continuous distribution of $\Xd_{k}$, the resultant procedure converges. Furthermore, the error associated with this procedure is bounded by a constant, which is dependent on the parameters used. As a result, the integral in \eqref{Eq: Integrated Distortion} may be rewritten as a sum over the codewords in quantizer $\Gamma_{k}$ and their associated probabilities.

Then an approximate value for the distortion may be computed as
\[
    \widetilde{D}\big(\Gamma_{k+1}\big)\approx D\big(\Gamma_{k+1}\big):=\sum_{i=1}^{N_k}\E{\big\|\U(\cwt{k}{i},Z_{k+1})-\Xq_{k+1}\big\|^2}\P\big(\Xq_k=\cwt{k}{i}\big).
\]
Here, $N_k$ is the cardinality of the quantizer $\Gamma_k$ at time step $k$, which is allowed to vary. With this definition of $D\big(\Gamma_{k+1}\big)$ we may now specify a Newton-Raphson iteration in order to compute the quantizer at time step $k+1$, which minimizes the distortion.

Given the quantizer at time $t_k$, represented as a column vector $\mGamma_{k}$, and the associated probabilities, $\P\big(\Xq_k=\cwt{k}{i}\big)$ for $1\leq i\leq N_k$, the Newton-Raphson iteration for the quantizer $\mGamma_{k+1}$, at time $t_{k+1}$, is given by
\begin{equation}
    \mGamma_{k+1}^{(n+1)} = \mGamma_{k+1}^{(n)}-\left[\nabla^2 D\left(\mGamma_{k+1}^{(n)}\right)\right]^{-1}\nabla D\left(\mGamma_{k+1}^{(n)}\right),\label{Eq: RMQ Newton Iteration}
\end{equation}
for $0\leq n<n_{\mathrm{max}}$.

\begin{figure}[t!]
    \begin{center}
        \includegraphics[width=0.7\columnwidth]{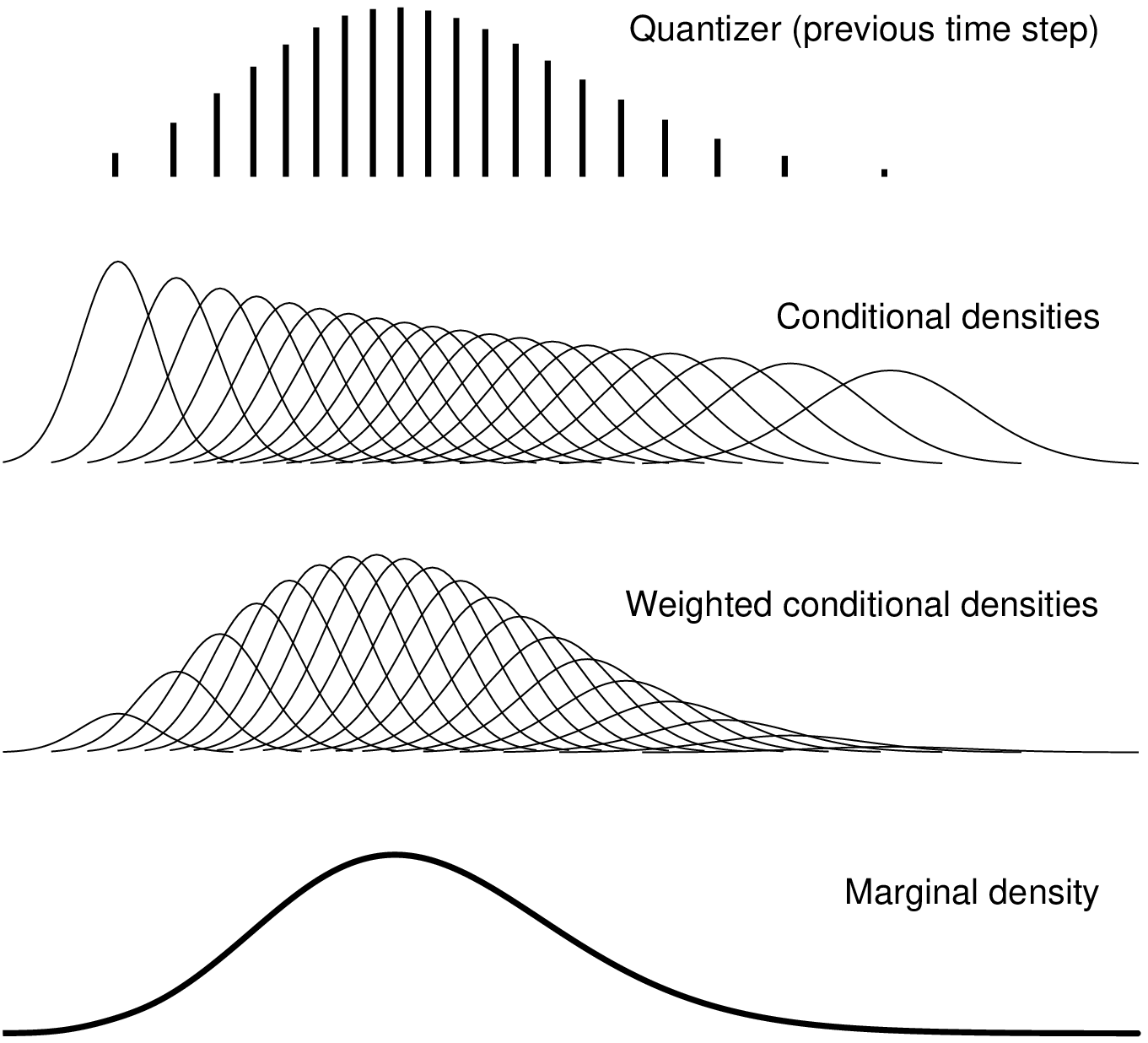}
    \end{center}
    \caption{Illustration of the RMQ algorithm.}
    \label{Fig: RMQGraphic}
\end{figure}

Before developing the mathematics further, we pause to provide an intuitive explanation of how the RMQ algorithm proceeds. Figure \ref{Fig: RMQGraphic} is a depiction of the process that occurs. The top panel shows the quantizer at time step $k$. Conditional on each codeword, a Gaussian Euler update is propagated (second panel). In panel three, these updates are weighted by the probability of the associated originating codeword and summed to produce the marginal density at time step $k+1$, as shown in the final panel. The distribution associated with this marginal density is the distribution that is quantized to produce the quantizer at time step $k+1$. This process is repeated until the quantizer at the final time is produced.

We now proceed to derive the quantities required for the Newton-Raphson iterations. To summarize notation, we write the update in affine form as
\begin{equation}
    \U(\cwt{k}{i},Z_{k+1})=:U_{k+1}^i=\affm{k}{i}Z_{k+1}^i+\affc{k}{i},\label{Eq: Affine Form}
\end{equation}
where
\[
    \affm{k}{i}:=b(\cwt{k}{i})\sqrt{\Delt}\qquad\text{and}\qquad
    \affc{k}{i}:=\cwt{k}{i}+a(\cwt{k}{i})\Delt,
\]
with $Z_{k+1}^i\sim\mathscr{N}(0,1)$ identically distributed to $Z_{k+1}$. Here, we have introduced a new index $i$ for the random variable $Z^i_{k+1}$ anticipating that it may depend on $\cwt{k}{i}$. This is redundant in the case of the Euler update because $Z_{k+1}$ is a standard normal random variate irrespective of starting point. This more general notation will become necessary when we analyse more general cases (see Section \ref{Sec: Higher Order}). We denote the corresponding density and distribution functions by $f_{Z_{k+1}^i}$ and $F_{Z_{k+1}^i}$ respectively.

With this notation in place, the approximate marginal distribution of $\Xd_{k+1}$ is
\begin{align}
    F_{\Xd_{k+1}}(x)&=\int_{\R}\P(\U(y,Z_{k+1})\leq x)\,d\P\big(\Xd_k\leq y\big)\notag\\
    &\approx\sum_{i=1}^{N_k}\P(U_{k+1}^i\leq x)\P\big(\Xq_k=\cwt{k}{i}\big)\notag\\
    &=\sum_{i=1}^{N_k}\left[H(-\affm{k}{i})+\sgn(\affm{k}{i})F_{Z_{k+1}^i}\!\!\left(\frac{x-\affc{k}{i}}{\affm{k}{i}}\right)\right]\P\big(\Xq_k=\cwt{k}{i}\big), \label{Eq: Implied Marginal Distribution}
\end{align}
where $H(\cdot)$ is the Heaviside step function and $\sgn(\cdot)$ is the signum function. The last step follows due to the fact that the left-hand probability on the penultimate line may be written as
\[
    \P(U_{k+1}^i\leq x)=
    \begin{cases}
        \P\left(Z_{k+1}^i\leq \frac{x-\affc{k}{i}}{\affm{k}{i}}\right) & \text{for $\affm{k}{i}\geq 0$}\\[3mm]
        1-\P\left(Z_{k+1}^i\leq \frac{x-\affc{k}{i}}{\affm{k}{i}}\right) & \text{for $\affm{k}{i}<0$.}
    \end{cases}
\]
It should be noted that, for the Euler update, $\affm{k}{i}$ is a proxy for the volatility of the SDE and its positivity is usually guaranteed, in which case \eqref{Eq: Implied Marginal Distribution} may be simplified. However, we persist with this formulation because in the general case $\affm{k}{i}$ may not be guaranteed to be positive.

The elements of the gradient of the distortion $\nabla D(\Gamma_{k+1})$ may then be written as
\begin{align}
    \frac{\partial D\left(\Gamma_{k+1}\right)}{\partial\cwt{k+1}{j}}
    &=2\sum_{i=1}^{N_k}
    \Ebig{\ind{U_{k+1}^i\in R_j(\Gamma_{k+1})}
    \left(\cwt{k+1}{j}-U_{k+1}^i\right)}\P\big(\Xq_k=\cwt{k}{i}\big)\notag\\
    &=2\sum_{i=1}^{N_k}
    \int_{U_{k+1}^i\in R_j(\Gamma_{k+1})}
    \left(\cwt{k+1}{j}-U_{k+1}^i\right)d\P(Z_{k+1}^i\leq x)\P\big(\Xq_k=\cwt{k}{i}\big),\label{Eq: Gradient Elements}
\end{align}
where $1\leq j\leq N_{k+1}$ is the index tracking the elements of the $t_{k+1}$ quantizer, and the index associated with the $t_k$ quantizer is $i$. The integration bounds in \eqref{Eq: Gradient Elements} must now be expressed in terms of the variable of integration.

Here, as in Section \ref{Sec: Vector Quantization}, $U_{k+1}^i\in R_j(\Gamma_{k+1})$ is equivalent to the inequality
\begin{equation}
    \rm{j}_{k+1}<U_{k+1}^i\leq\rp{j}_{k+1}\qquad\text{with}\qquad r_{k+1}^{j\pm}=\tfrac{1}{2}(\cwt{k+1}{j\pm1}+\cwt{k+1}{j}),\label{Eq: RMQ Regions}
\end{equation}
and $\rm{1}_{k+1}=-\infty$ and $\rp{N_{k+1}}_{k+1}=\infty$ by definition. Defining the conditionally normalized region boundaries,
\[
    r_{k+1}^{i,j\pm}=\frac{r_{k+1}^{j\pm}-\affc{k}{i}}{\affm{k}{i}},
\]
allows the inequality to be written in terms of the random variable $Z^i_{k+1}$ as
\[
    U_{k+1}^i\in R_j(\Gamma_{k+1})=
    \begin{cases}
        \Rm{k+1}{i,j}<Z^i_{k+1}\leq\Rp{k+1}{i,j} & \text{for $\affm{k}{i}\geq 0$}\\[3mm]
        \Rm{k+1}{i,j}>Z^i_{k+1}\geq\Rp{k+1}{i,j} & \text{for $\affm{k}{i}<0$,}
    \end{cases}
\]
which can now be used as the range over which the integration is taken.

Directly evaluating \eqref{Eq: Gradient Elements}, each element of the gradient of the distortion at time $t_{k+1}$ is given by
\begin{align*}
    \frac{\partial D(\Gamma_{k+1})}{\partial\cwt{k+1}{j}}
    &=2\sum_{i=1}^{N_k}\left[(\cwt{k+1}{j}-\affc{k}{i})\sgn(\affm{k}{i})\left(F_{Z_{k+1}^i}(\Rp{k+1}{i,j})-F_{Z_{k+1}^i}(\Rm{k+1}{i,j})\right)\right.\\
    &\qquad\qquad\left.-|\affm{k}{i}|\left(M_{Z_{k+1}^i}^1(\Rp{k+1}{i,j})-M_{Z_{k+1}^i}^1(\Rm{k+1}{i,j})\right)\right]\P\big(\Xq_k=\cwt{k}{i}\big).\\
\intertext{Furthermore, the diagonal of the tridiagonal Hessian, $\nabla^2D(\Gamma_{k+1})$, is given by}
    \frac{\partial^2D(\Gamma_{k+1})}{\partial \big(\cwt{k+1}{j}\big)^2}&=\sum_{i=1}^{N_k}\bigg[2\sgn(\affm{k}{i})\left(F_{Z_{k+1}^i}(\Rp{k+1}{i,j})-F_{Z_{k+1}^i}(\Rm{k+1}{i,j})\right)\\
    &\qquad\qquad+\frac{1}{2|\affm{k}{i}|}f_{Z_{k+1}^i}(\Rp{k+1}{i,j})(\cwt{k+1}{j}-\cwt{k+1}{j+1})\\
    &\qquad\qquad+\frac{1}{2|\affm{k}{i}|}f_{Z_{k+1}^i}(\Rm{k+1}{i,j})(\cwt{k+1}{j-1}-\cwt{k+1}{j})\bigg]\P\big(\Xq_k=\cwt{k}{i}\big),\\
\intertext{with the super-diagonal and sub-diagonal elements given by}
    \frac{\partial^2D(\Gamma_{k+1})}{\partial \cwt{k+1}{j}\partial \cwt{k+1}{j+1}}&=\sum_{i=1}^{N_k}\frac{1}{2|\affm{k}{i}|}f_{Z_{k+1}^i}(\Rp{k+1}{i,j})(\cwt{k+1}{j}-\cwt{k+1}{j+1})\P\big(\Xq_k=\cwt{k}{i}\big)\\
\intertext{and}
    \frac{\partial^2D(\Gamma_{k+1})}{\partial \cwt{k+1}{j}\partial \cwt{k+1}{j-1}}&=\sum_{i=1}^{N_k}\frac{1}{2|\affm{k}{i}|}f_{Z_{k+1}^i}(\Rm{k+1}{i,j})(\cwt{k+1}{j-1}-\cwt{k+1}{j})\P\big(\Xq_k=\cwt{k}{i}\big),
\end{align*}
respectively.

Although these expressions may appear complex, they are simply summations over the density function, cumulative distribution function and first lower partial expectation of the random variable, $Z^i_{k+1}$, and are thus easy to compute when these functions are known.

All the detail required to implement the Newton iteration \eqref{Eq: RMQ Newton Iteration} has now been provided with the exception of the initial guess. In all applications considered in this paper, we have assumed that $N_k=N$ for $1\leq k\leq K$, and used the quantizer from the previous time step as the initial guess, i.e., $\mGamma_{k+1}^{(0)}=\mGamma_{k}$.

\subsection{Efficient Implementation}
\label{Sec: Efficient Implementation}
As in Section \ref{Sec: VQ Efficient Implementation}, where we provided an efficient matrix formulation for the Newton-Raphson iteration required for VQ, RMQ is also amenable to a matrix specification. This aids simple and computationally efficient implementation.

Aside from a guess for $\mGamma_{k+1}$, we require the following time-indexed column vectors
\begin{align*}
    [\mm_k]_i&=\affm{k}{i},& \qquad
    [\mc_k]_i&=\affc{k}{i},& & & &1\leq i\leq N_k\\
\intertext{and}
    [\Delta\mGamma_{k+1}]_i&=\cwt{k+1}{i+1}-\cwt{k+1}{i}, & & & & & &1\leq i\leq N_{k+1}-1.
\end{align*}
The row vector of probabilities
\[
    \mp_k=[\P(\Xq_k=\cwt{k}{1}),\dots,\P(\Xq_k=\cwt{k}{N_k})],
\]
is retained and a row-vector of ones of length $d$ is denoted by $\mJ_d.$ With the exception of $\Delta\mGamma_{k+1}$, which must be recomputed before each Newton-Raphson iteration, the other vectors are computed once per time step.

Before each Newton-Raphson iteration, three matrices must be computed in terms of the new estimate of $\mGamma_{k+1}$: an $N_{k}\times N_{k+1}$ matrix of transition probabilities
\begin{align}
    [\mP_{k+1}]_{i,\,j}&=\P(\Xq_{k+1}=\cwt{k+1}{j}|\Xq_k=\cwt{k}{i})\notag\\
    &=\sgn(\affm{k}{i})\left[F_{Z^i_{k+1}}(\Rp{k+1}{i,j})-F_{Z^i_{k+1}}(\Rm{k+1}{i,j})\right],\notag
\intertext{another matrix, of the same size, of lower partial moment values}
    [\mM_{k+1}]_{i,j}&=M_{Z^i_{k+1}}^1(\Rp{k+1}{i,j})-M_{Z^i_{k+1}}^1(\Rm{k+1}{i,j})
    \label{Eq: M values}
\intertext{and an $N_k\times N_{k+1}-1$ matrix of density values at the positive region boundaries}
    [\mf_{k+1}]_{i,j}&=f_{Z^i_{k+1}}(\Rp{k+1}{i,j}).\notag
\end{align}
The gradient of the distortion function at time step $k+1$ may then be written in terms of these vectors and matrices as
\begin{equation}
    \nabla D\left(\mGamma_{k+1}\right)^\top= 2\mp_k\left(\big((\mGamma_{k+1}\mJ_{N_k})^\top-\mc_k\mJ_{N_{k+1}}\big)\hprod \mP_{k+1}-(|\mm_k|\mJ_{N_{k+1}})\hprod \mM_{k+1}\right), \label{Eq: RMQ Matrix Distortion Gradient}
\end{equation}
where $\hprod$ is the Hadamard (or element-wise) product.

The super and sub-diagonal elements of the (tridiagonal) Hessian matrix, $\nabla^2 D\left(\mGamma_{k+1}\right)$, are given by the vector
\begin{equation}
    \mho=-\tfrac12\mp_k\left(\left(\lvert\mm_k\rvert^{\hprod-1}\mJ_{(N_{k+1}-1)}\right)\hprod
    \mf_{k+1}\hprod(\Delta\mGamma_{k+1}\mJ_{N_k})^\top\right), \label{Eq: Off-Diagonals}
\end{equation}
while the main diagonal is given by
\begin{equation}
    \mhm=2\mp_k\mP_{k+1}+\big[\mho|0\big]+\big[0|\mho\big],\label{Eq: Main Diagonal}
\end{equation}
where ${\hprod-1}$ in the exponent refers to the element-wise inverse.

Equations \eqref{Eq: RMQ Matrix Distortion Gradient}, \eqref{Eq: Off-Diagonals} and \eqref{Eq: Main Diagonal} provide the necessary components required for implementation of the Newton-Raphson iteration in \eqref{Eq: RMQ Newton Iteration}. After the requisite number of iterations, the probabilities associated with the final quantizer $\mGamma_{k+1}$ are computed using
\[
    \mp_{k+1}=\mp_k\mP_{k+1},
\]
where $\mP_{k+1}$ must be recomputed in terms of the final $\mGamma_{k+1}$. Thus, the matrix formulation presented here allows RMQ to be interpreted as the propagation of an inhomogeneous, discrete time Markov chain, where $\mGamma_{k}$ represents the Markov states at time step $k$, the probability of being in those states is $\mp_k$, and the associated transition probability matrix is $\mP_{k+1}$. Sometimes in the literature the transition probability matrix between time step $k$ and $k+1$ is represented as $\mP_{k,k+1}$, we have chosen to omit the first index.

\subsection{Example}
\label{Sec: RMQ Example}

\begin{figure}[t!]
     \begin{center}
         \includegraphics[width=\columnwidth]{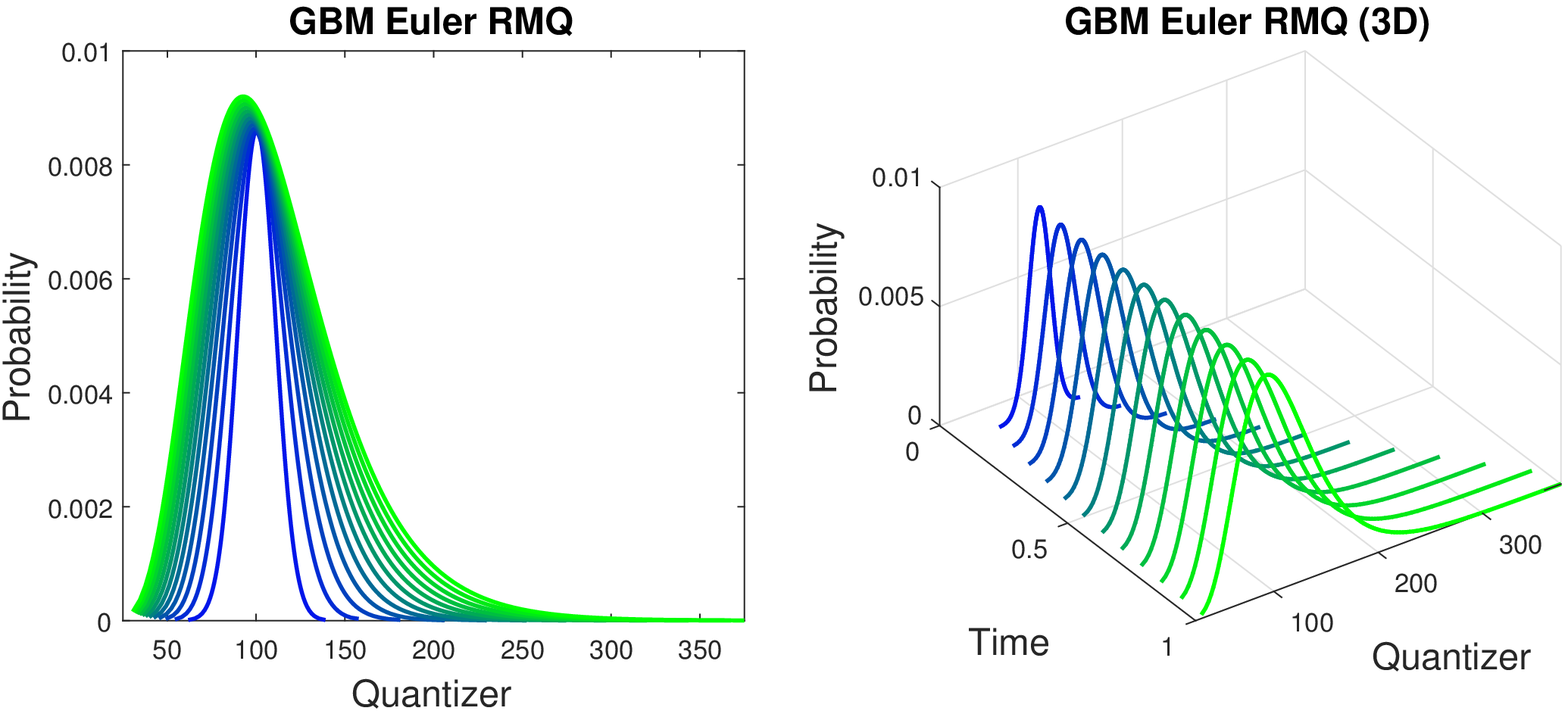}
     \end{center}
     \caption{Time evolution of quantizers for the GBM process.}
     \label{Fig: Plot_RMQ_GBM}
\end{figure}

As a first example of the RMQ algorithm, Figure \ref{Fig: Plot_RMQ_GBM} shows the evolution of the quantizers through time for the canonical (risk-neutral) geometric Brownian motion process
\begin{equation}
    dS_t=rS_t\,dt+\sigma S_t\,dW_t,\label{Eq: GBM SDE}
\end{equation}
using parameters $S_0=100$, $r=5\%$ and $\sigma=30\%$. The RMQ parameters used were $T=1$, $K=12$,  $\Delta t=T/K$ and $N_k=200$ for all $k$, with $n_{\mathrm{max}}=50$ for the VQ algorithm and $n_{\mathrm{max}}=5$ for the RMQ algorithm. Unless otherwise stated, these parameters are used wherever geometric Brownian motion is used for other calculations in the paper. In these plots, the colour of the quantizer indicates the associated time step, with lines in blue closer to initial time and lines in green closer to final time. This convention is kept throughout.

%
%

%% file: Sections/HigherOrder.tex
Given the general formulation of the RMQ algorithm in the previous section, we now explore two high-order extensions: the Milstein scheme and a simplified weak order 2.0 scheme. Any numerical scheme for an SDE that can be written in the affine form \eqref{Eq: Affine Form}, may be used with the RMQ algorithm as long as the CDF, PDF and lower partial expectation of the random variable $Z^i_{k+1}$ can be computed for all $i$ and $k$.

\subsection{The Milstein Scheme}

The \cite{mil1975approximate} scheme for the SDE in \eqref{Eq: SDE} is given by
\[
    \Xd_{k+1}=\Xd_{k}+a(\Xd_k)\Delt+b(\Xd_k)\sqrt{\Delt}Z_{k+1}+\tfrac12b(\Xd_k)\dx{b}{\Xd_k}\Delt\left( Z_{k+1}^2-\Delt\right),
\]
for $0\leq k<K$, where $\Delt=T/K$ and $Z_{k+1}\sim\mathscr{N}(0,1)$ with initial value $\Xd_0=x_0$.
By completion of the square, this may be written as
\begin{align*}
    \Xd_{k+1}
    &=\Xd_{k}+\left(a(\Xd_{k})-\tfrac12b(\Xd_{k})\dx{b}{\Xd_{k}}\right)\Delt-\tfrac12b(\Xd_{k})\dx{b}{\Xd_{k}}^{-1}\\
    &\qquad+\tfrac12b(\Xd_{k})\dx{b}{\Xd_{k}}\Delt\left(Z_{k+1}+\left(\sqrt{\Delt}\dx{b}{\Xd_{k}}\right)^{-1}\right)^2.
\end{align*}
Thus, the update \eqref{Eq: Affine Form} may be written in the affine form required as
\begin{align*}
    U_{k+1}^i&=\affm{k}{i}Z_{k+1}^i+\affc{k}{i},\\
\intertext{where}
	\affm{k}{i}&=\tfrac12b(\cwt{k}{i})\dx{b}{\cwt{k}{i}}\Delt
\intertext{and}
    \affc{k}{i}&=\cwt{k}{i}+\left(a(\cwt{k}{i})-\tfrac12b(\cwt{k}{i})\dx{b}{\cwt{k}{i}}\right)\Delt-\tfrac12b(\cwt{k}{i})\dx{b}{\cwt{k}{i}}^{-1}.
\end{align*}
The random variable $Z_{k+1}^i$ is now noncentral chi-squared distributed with one degree of freedom and noncentrality parameter
\[
    \lambda_{k+1}^i=\left(\sqrt{\Delt}\dx{b}{\cwt{k}{i}}\right)^{-2}.
\]
It is important to note that, unlike the Euler-Maruyama case, the distribution of the random variable $Z^i_{k+1}\sim\chi^{\prime2}(1,\lambda_{k+1}^i)$ now depends on the codeword $\cwt{k}{i}$.

Although the Milstein scheme possesses a strong order of convergence of $1$, compared to the Euler scheme, which only has strong order of convergence of $\frac{1}{2}$, both schemes have a weak order of convergence of $1$.
Thus, while the Milstein scheme is more accurate in a strong sense than the Euler scheme, we require a different update for higher weak order convergence, which is needed when approximating expectations of financial payoffs. We now explore such a scheme.

\subsection{A Weak Order \texorpdfstring{$2.0$}{2.0} Taylor Scheme}

While it is not possible to write a weak order 2.0 Taylor scheme in the affine form required, the simplified weak order 2.0 scheme of \cite{kloeden1999numerical} is amenable. This scheme is given by
\begin{align*}
	\Xd_{k+1}&=\Xd_k+a(\Xd_k)\Delt+b(\Xd_k)\sqrt{\Delt}Z_{k+1}
    +\tfrac{1}{2}b(\Xd_k)\dx{b}{\Xd_k}\Delt(Z_{k+1}^2-1)\\
	&\qquad+\tfrac{1}{2}\left(\dx{a}{\Xd_k}b(\Xd_k)+a(\Xd_k)\dx{b}{\Xd_k}+ \tfrac{1}{2}\ddx{b}{\Xd_k}b^2(\Xd_k) \right)(\Delt)^{\frac{3}{2}}Z_{k+1}\\
	&\qquad+\tfrac{1}{2}\left(a(\Xd_k)\dx{a}{\Xd_k} + \tfrac{1}{2}\ddx{a}{\Xd_k}b^2(\Xd_k)\right)(\Delt)^2,
\end{align*}
for $0\leq k<K$, where $\Delt=T/K$ and $Z_{k+1}\sim\mathscr{N}(0,1)$ with initial value $\Xd_0=x_0$. Again, completion of squares is used to write this update in the required affine form,
\begin{align*}
    U_{k+1}^i&=\affm{k}{i}Z_{k+1}^i+\affc{k}{i},\\
\intertext{where}
	\affm{k}{i}&=\tfrac{1}{2}b(\cwt{k}{i})\dx{b}{\cwt{k}{i}}\Delt\\
\intertext{and}
    \affc{k}{i}&=\cwt{k}{i}+\left(a(\cwt{k}{i})-\tfrac12b(\cwt{k}{i})\dx{b}{\cwt{k}{i}}\right)\Delt
    +\tfrac{1}{2}\left(a(\cwt{k}{i})\dx{a}{\cwt{k}{i}} + \tfrac{1}{2}\ddx{a}{\cwt{k}{i}}b^2(\cwt{k}{i})\right)(\Delt)^2\\
    &\qquad-\frac{\left(b(\cwt{k}{i})+\frac{1}{2}\left(\dx{a}{\cwt{k}{i}}b(\cwt{k}{i}) + a(\cwt{k}{i})\dx{b}{\cwt{k}{i}} + \frac{1}{2}\ddx{b}{\cwt{k}{i}}b^2(\cwt{k}{i})  \right)\Delt \right)^2}{2b(\cwt{k}{i})\dx{b}{\cwt{k}{i}}}.
\end{align*}
Here, $Z_{k+1}^i$ is again noncentral chi-squared distributed with one degree of freedom, with noncentrality parameter given by
\begin{equation*}
	\lambda^i_{k+1} = \left(\frac{b(\cwt{k}{i}) + \frac{1}{2}\left(\dx{a}{\cwt{k}{i}}b(\cwt{k}{i}) + a(\cwt{k}{i})\dx{b}{\cwt{k}{i}} + \frac{1}{2}\ddx{b}{\cwt{k}{i}}b^2(\cwt{k}{i}) \right)\Delt}{b(\cwt{k}{i})\dx{b}{\cwt{k}{i}} \sqrt{(\Delt)}}\right)^2,
\end{equation*}
or, more succinctly, $Z^i_{k+1}\sim\chi^{\prime2}(1,\lambda_{k+1}^i)$.

\subsection{Examples}
\label{Sec: Higher Order Examples}

To illustrate the accuracy of the above schemes, the RMQ algorithm is applied to geometric Brownian motion, as previously described by \eqref{Eq: GBM SDE}, and the constant elasticity of variance (CEV) process. The SDE for the CEV process is
\[
    dS_t=rS_t\,dt+\sigma S_t^\alpha\,dW_t,
\]
and, in the examples that follow, the process-specific parameters chosen were
$S_0=100$, $r=5\%$ and $\alpha=0.7$ and $\sigma=\sigma_{\text{LN}}S_0^{1-\alpha}$, with $\sigma$ given in terms of the instantaneous log-normal volatility $\sigma_{\text{LN}}=30\%$. In the case of GBM the parameters in Section \ref{Sec: RMQ Example} were used. For both GBM and CEV, the RMQ specific parameters mentioned in that section were also used, unless stated otherwise.

\begin{figure}[t!]
     \begin{center}
         \includegraphics[width=\columnwidth]{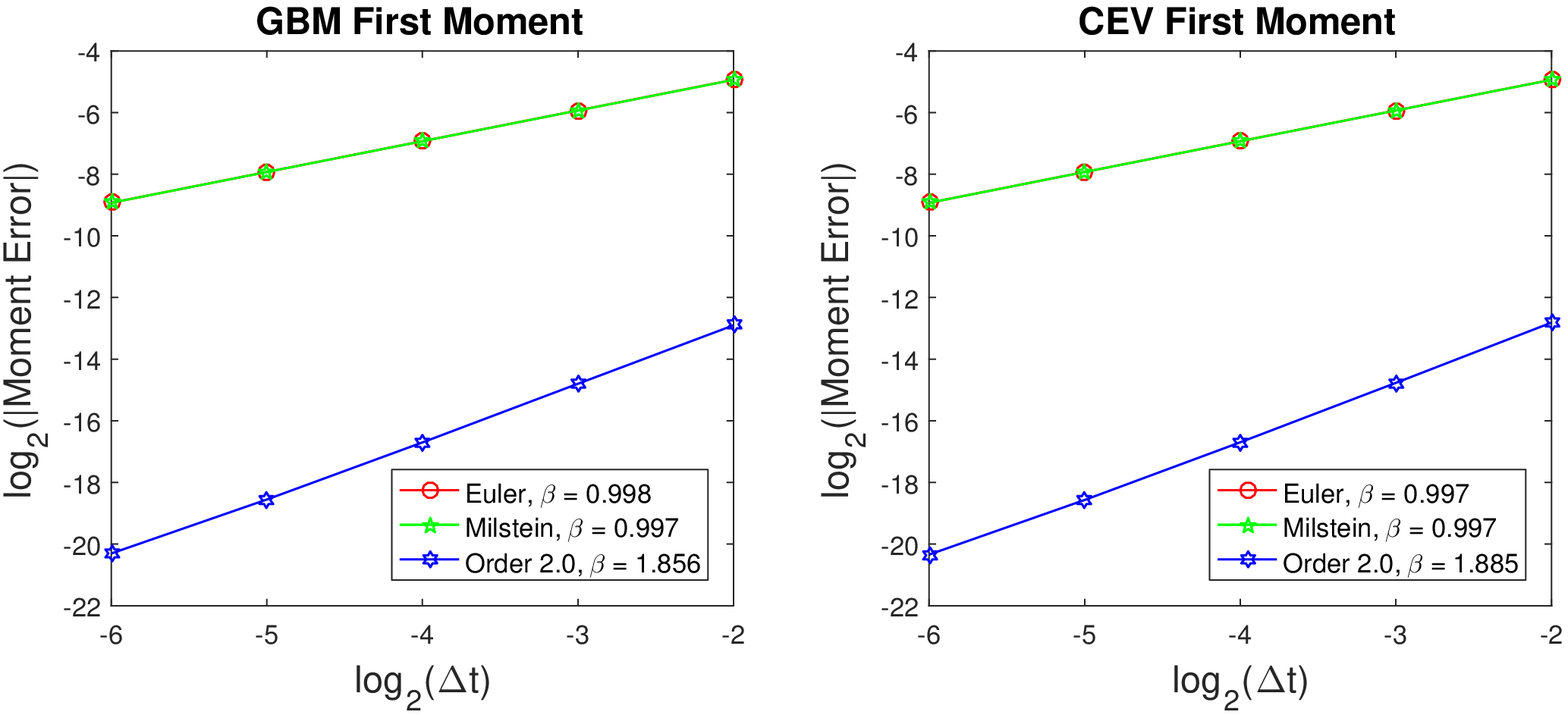}
     \end{center}
     \caption{Convergence of the first moment for GBM and CEV.}
     \label{Fig: Plot_weakError_GBM}
\end{figure}

In Figure \ref{Fig: Plot_weakError_GBM} we show numerical evidence for weak order convergence. The absolute value of the difference between the first moment of the resulting terminal quantizer and the true moment is plotted for a range of time step sizes. In each case, the horizontal axis provides the base-2 logarithm of the step size and the vertical axis the base-2 logarithm of the error in the first moment. Thus, the slope of each graph reflects the power of the step size and hence the order of weak convergence for the error. In the figure legends, the regressed gradients of the graphs, denoted by $\beta$, indicate the weak orders of convergence. As is theoretically expected, see \cite{kloeden1999numerical}, both the Euler and Milstein schemes have approximately weak order one convergence, whereas the simplified weak order 2.0 scheme reaches a weak order close to two. Therefore, the latter scheme is by magnitude more accurate and can be expected to produce results with substantially lower error than the Euler scheme when valuing contingent claims. For increased accuracy in these plots, the cardinality used was $N_k=1\,000$.

\begin{figure}[t!]
     \begin{center}
         \includegraphics[width=\columnwidth]{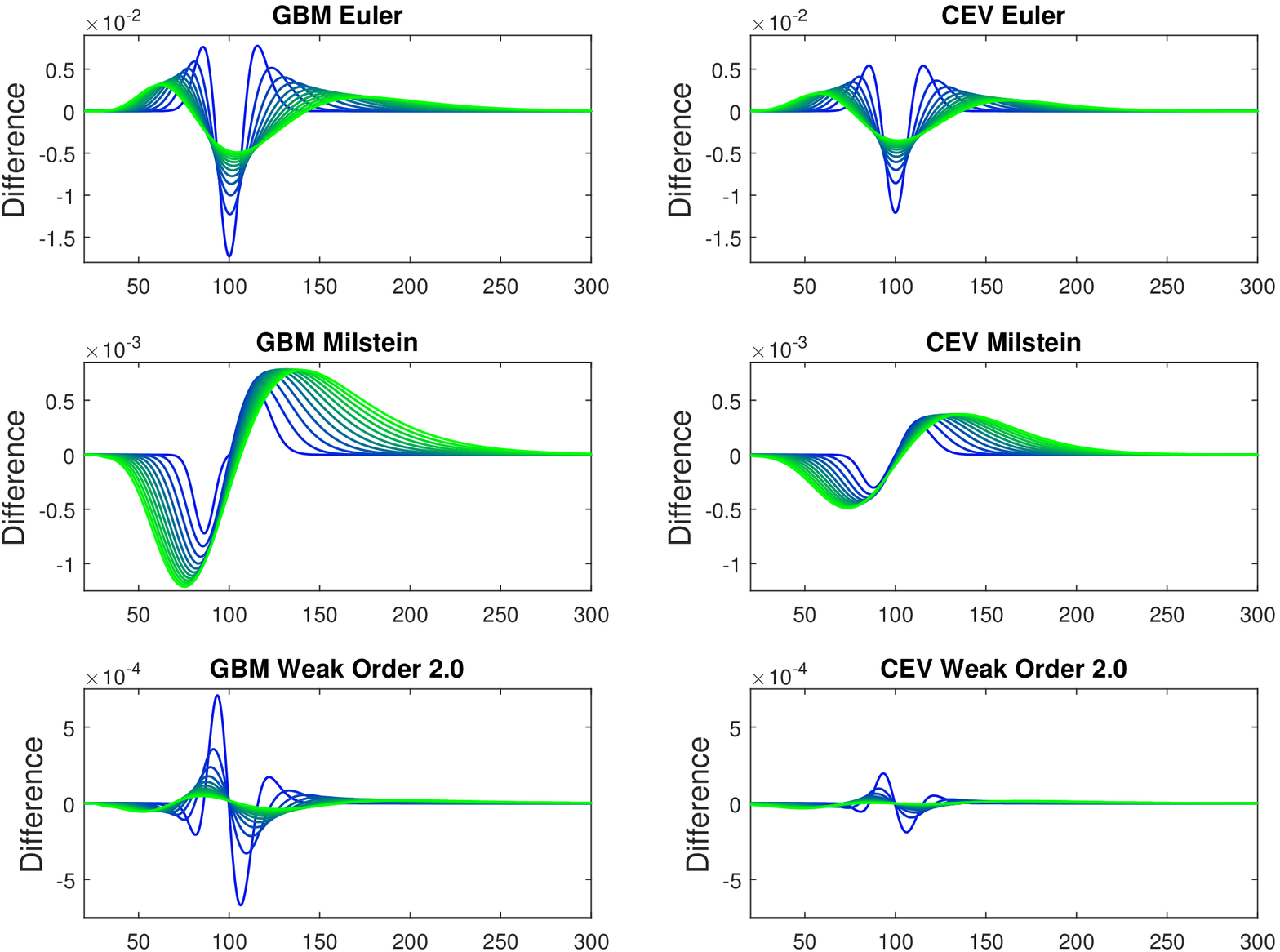}
     \end{center}
     \caption{The error in the marginal distribution implied by quantization for GBM and CEV.}
     \label{Fig: Plot_Densities}
 \end{figure}

Since the true conditional distributions for GBM and CEV are known in closed form, we can compare the approximate marginal distributions at time step $k+1$, as implied by the quantizer at time step $k$ and computed using \eqref{Eq: Implied Marginal Distribution}, with the exact distributions at time step $k+1$. In the case of the CEV process we used the analytical expression for the distribution given by \cite{lindsay2012simulation} adjusted for the drift component in the SDE.

For each of the three schemes, under the GBM and CEV test cases, the difference between the exact marginal distribution and the implied marginal distribution is plotted in Figure \ref{Fig: Plot_Densities}. Here we reverted to using a cardinality of $N_k=200$. Note the scale of the $y$-axes of the graphs in the figure --- from top to bottom, the magnitude of the error decreases by an order of magnitude in each successive row. This gives an indication of the improvement that can be expected when these higher order schemes are used to price contingent claims.


%% file: Sections/AbsorptionReflection.tex
Sometimes discrete-time approximations of an SDE may exhibit behaviour that is inconsistent with the true solution. For example, an Euler-Maruyama approximation of geometric Brownian motion or the CEV process can, under certain circumstances, generate negative values, even though the SDE specification guarantees non-negativity in each case. As a result, discrete-time Monte Carlo simulations are often modified to generate reflecting or absorbing behaviour at zero, see for example \cite{lord2010comparison}. In this section, we describe how the RMQ algorithm may be modified in a similar manner.

\subsection{Absorbing Boundary}

To model an absorbing boundary, the domain of the approximate marginal distribution of $\Xd$, see \eqref{Eq: Implied Marginal Distribution}, must be left-truncated at zero. Implementing the RMQ algorithm with a left limit of zero results in a quantizer that has probabilities at each time step that do not sum to unity. The probability that is not accounted for as a result of the domain truncation is the mass accumulated at the absorbing zero boundary. To compensate for this, the quantizer at each time step can be augmented with an extra codeword, which has a value of zero and a probability equal to one minus the sum of the probabilities associated with all the other codewords at that time step. The transition probability matrix may also be augmented in a consistent manner by realising that once the process attains the zero state it must remain in that state indefinitely, i.e., the conditional probability of moving from the absorbing state to any other state is zero, and, correspondingly, the conditional probability of remaining in the absorbing state is one.

Modifying the algorithm is straightforward and incurs no additional computational burden. Given that the elements of the previous quantizer $\mGamma_k$ are all positive, the affine form of the update \eqref{Eq: Affine Form}, will be negative when
\begin{align*}
	Z_{k+1}^i &<-\frac{ \affc{k}{i}}{\affm{k}{i}}.
\end{align*}
This implies that the domain of each $Z_{k+1}^i$ must be left-truncated at $-\tfrac{ \affc{k}{i}}{\affm{k}{i}}$ to ensure only positive codewords at time-step $k+1$. This is achieved by setting
\begin{equation}
	\Rm{k+1}{i,1} = -\frac{\affc{k}{i}}{\affm{k}{i}}, \label{Eq: Lower Bound}
\end{equation}
for $1\leq i\leq N_k$, in the implementation described in Section \ref{Sec: Efficient Implementation}. This is equivalent to assuming $\rm{1}_{k+1}=0$ in \eqref{Eq: RMQ Regions}. The rest of the algorithm proceeds without modification.

Of course, this all depends on the fact that the quantizer at the first time step, $\mGamma_1$, also has positive elements. This is achieved using an analogous truncation in the vector quantization algorithm. The initial guess for the Newton iteration must also ensure positivity.

\subsection{Reflecting Boundary}

\begin{figure}
     \begin{center}
         \includegraphics[width=0.7\columnwidth]{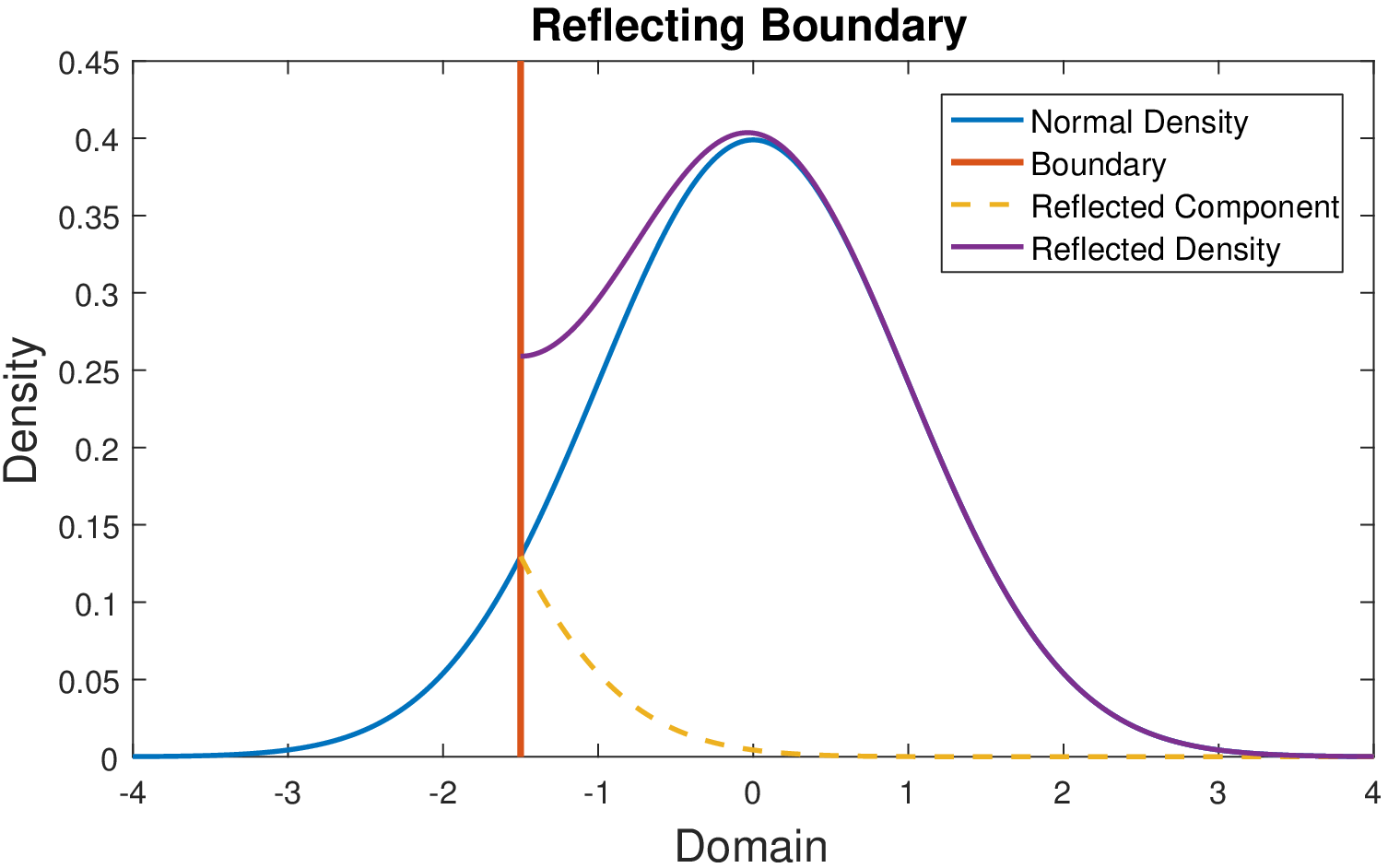}
     \end{center}
     \caption{Illustration of the standard Gaussian density reflected around $-1.5$.}
     \label{Fig: reflection}
\end{figure}
Figure \ref{Fig: reflection} shows $f(x)$, a density function  --- in this case a standard Gaussian density. The red line represents a reflecting boundary at $\bar{x}=-1.5$. The $f$ values to the left of the boundary are reflected and depicted by the dashed yellow line in the figure. These values are given by $f(2\bar{x}-x)$ for $x>\bar{x}$.

Thus, restricting the domain to $[\bar{x},\infty)$ the reflected density, denoted $\bar{f}(x)$, is given as the sum
\[
 	\bar{f}(x)=f(x)+f(2\bar{x}-x). 
\]
Direct integration of this expression over the integration limits from $\bar{x}$ to $x\in[\bar{x},\infty)$ gives the reflected distribution function
\[
    \bar{F}(x)=F(x)-F(2\bar{x}-x)
\]
and the first lower partial expectation function
\begin{equation}
    \bar{M}^1(x)=M^1(x)+M^1(2\bar{x}-x)-2\bar{x}F(2\bar{x}-x)-2M^1(\bar{x})+2\bar{x}F(\bar{x}), \label{Eq: Reflected M1}
\end{equation}
where $F(x)$ and $M^1(x)$ are the un-reflected distribution and first lower expectation functions associated with $f$. This may be applied, not only to the Gaussian case, but also to the noncentral chi-squared cases required for the higher order updates.

Modifying the RMQ algorithm to allow for a reflecting boundary at zero requires two changes to the implementation described in Section \ref{Sec: Efficient Implementation}. Firstly, the lower bound for the integration, i.e., the domain of the $Z_{k+1}^i$ random variable in each affine update, must be left-truncated by replacing the furthest left region boundary as in \eqref{Eq: Lower Bound} above. Secondly, the density, distribution and first lower partial expectation functions associated with each random variable, must be replaced by their reflected counterparts
\begin{align}
    \bar{f}_{Z_{k+1}^i}(x)&=f_{Z_{k+1}^i}(x)+f_{Z_{k+1}^i}(2\bar{x}^i_k-x),\notag\\
 	 \bar{F}_{Z_{k+1}^i}(x)&=F_{Z_{k+1}^i}(x)-F_{Z_{k+1}^i}(2\bar{x}^i_k-x),\notag\\
\intertext{and}
    \bar{M}_{Z_{k+1}^i}^1(x)&=M_{Z_{k+1}^i}^1(x)+M_{Z_{k+1}^i}^1(2\bar{x}^i_k-x)- 2\bar{x}^i_kF_{Z_{k+1}^i}(2\bar{x}^i_k-x), \label{Eq: Reflected M1 RMQ}
\end{align}
for $x\in[\bar{x}^i_k, \infty)$, where	 $\bar{x}^i_k=-\tfrac{\affc{k}{i}}{\affm{k}{i}}$. The remainder of the algorithm proceeds as normal. The astute reader will have noticed that there are two terms missing in \eqref{Eq: Reflected M1 RMQ} when compared with \eqref{Eq: Reflected M1}. The reason for this omission is that these terms are constants for each $i$ and that the RMQ algorithm always only requires differences of partial moment terms, as seen in \eqref{Eq: M values}. There is, therefore, a cancelation of the constant terms when this difference is taken and thus we may use a definition that excludes them.

As in the case of the absorbing boundary, the analogous reflection must be applied in the vector quantization algorithm to ensure that $\mGamma_1$ is consistent.

\subsection{Examples}

\begin{figure}[t!]
    \begin{center}
        \includegraphics[width=\columnwidth]{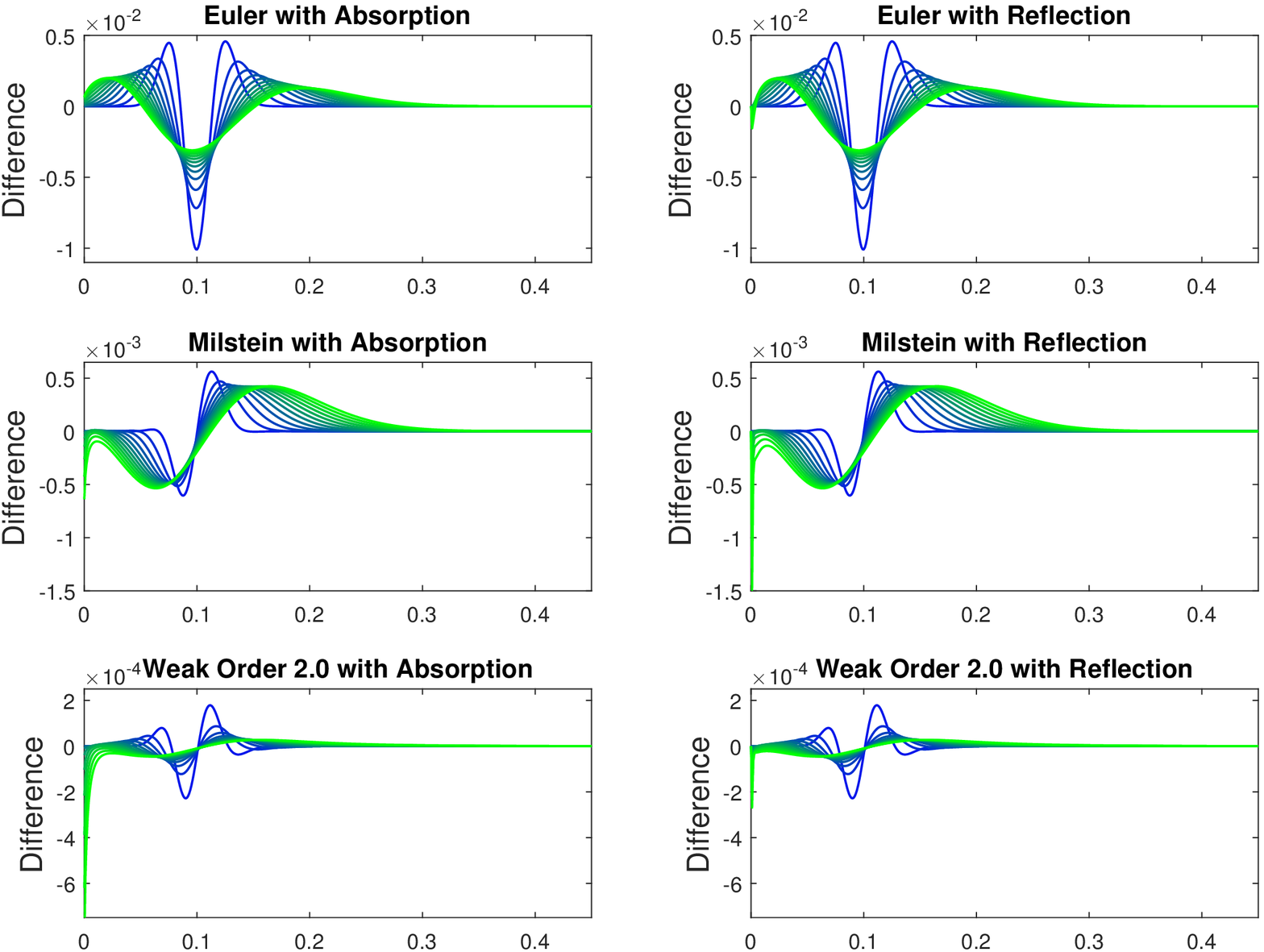}
    \end{center}
    \caption{The difference between the true and approximate marginal distributions for the CEV process. The left column shows the case of absorption while the right shows reflection.} \label{Fig: Plot_reflection_cev}
\end{figure}

It is well known that when $0<\alpha<0.5$, the CEV process may reach zero and that this state may be either absorbing or reflecting.  \cite{lindsay2012simulation} give the corresponding marginal distributions for both these cases (it is easy to adjust their formulations to account for the drift term in the SDE for the CEV process). In Section \ref{Sec: Higher Order Examples} we considered the CEV process with $\alpha>0.5$, which only allows absorption at zero. Now consider the case where $S_0=0.5$, $\alpha=0.35$ and $\sigma_{\text{LN}}=50\%$, with the rest of the parameters as before. Figure \ref{Fig: Plot_reflection_cev} shows the difference between the exact marginal distribution and the marginal distribution implied by RMQ for the three schemes as modified to account for an absorbing boundary (left) and a reflecting boundary (right). As before, the scale of the graphs changes from top to bottom, indicating the improvement as a result of the choice of the schemes.

Note that, under this choice of parameters for the CEV model, the standard RMQ formulation of Section \ref{Sec: RMQ} fails. Without implementing the modifications for either absorption or reflection proposed in this section, some codewords become negative at a certain point in the execution of the RMQ algorithm, leading to discrete-time updates with imaginary values.


%% file: Sections/Pricing.tex
In this section, contingent claims are priced using the RMQ algorithm and the three update schemes are compared for accuracy. The claims priced include European, Bermudan and discretely-monitored barrier options under the dynamics of both geometric Brownian motion (GBM) and its generalization, the constant elasticity of variance (CEV) model.

The GBM model and its parameters are described in Section \ref{Sec: RMQ Example}, whereas the specification for the CEV model can be found in Section \ref{Sec: Higher Order Examples}. These parameters are used for pricing throughout this section. As is implied by these specifications, the continuously compounded interest rate is assumed constant with a value $r=5\%$. All option maturities are one year and the RMQ algorithm is executed using $K=12$, i.e., using monthly steps, with constant cardinality of $N_k=200$ for all $k$.



\subsection{European Option Pricing}

\begin{figure}
     \begin{center}
         \includegraphics[width=\columnwidth]{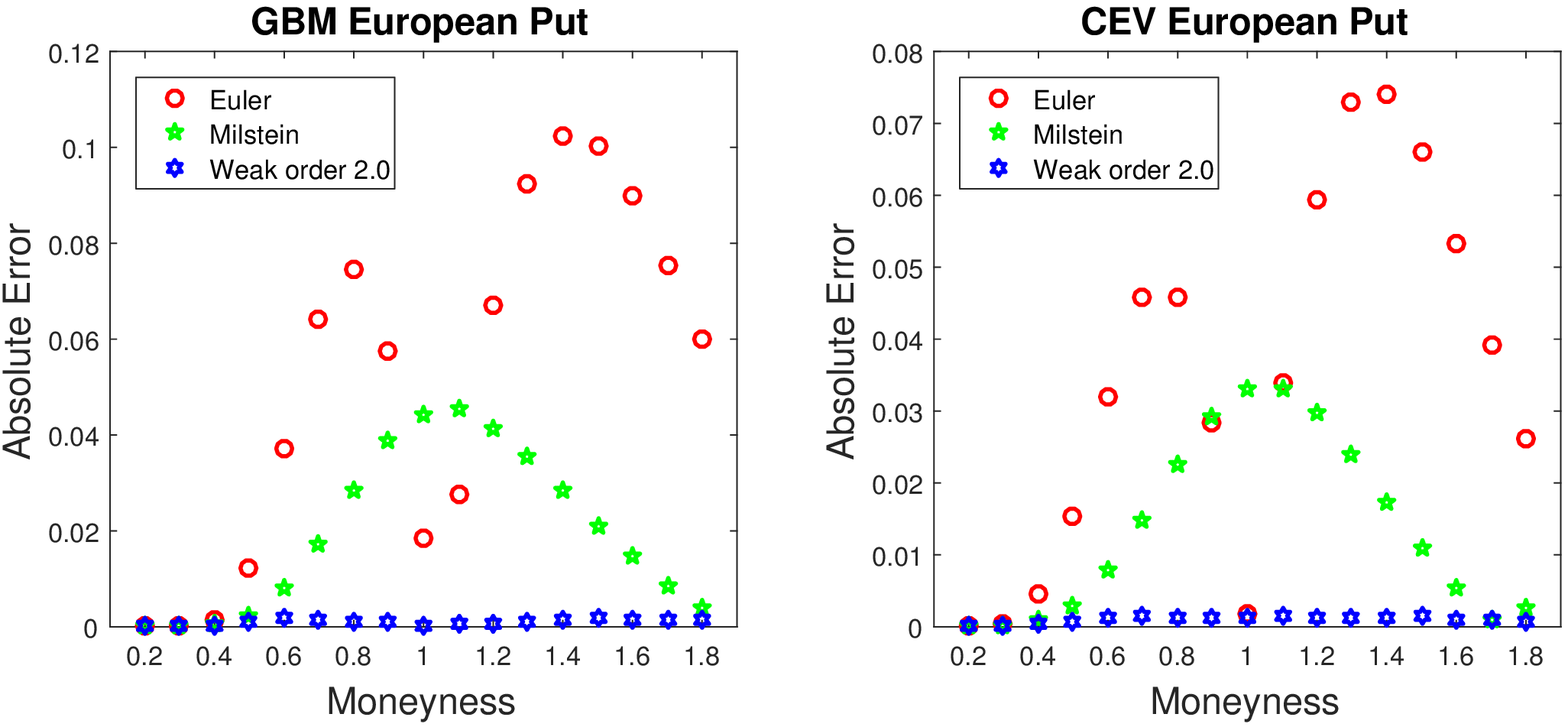}
     \end{center}
     \caption{Accuracy of GBM and CEV European put prices computed using RMQ, as compared to analytical solutions.}
     \label{Fig: European_Put}
 \end{figure}



Once a terminal quantizer has been obtained using the RMQ algorithm, a European option with payoff function $H(S,X)$ at maturity $T =t_K$, where $S$ represents the asset process and $X$ the strike, may be priced directly by using the expectation defined in \eqref{Eq: Expectation of a Functional}. The price is given by
\begin{equation}
	H_{0}= e^{-rT}\E{H(S_{T},X)}
	\approx e^{-rT}\mp_K H(\mGamma_K,X), \label{Eq: European Options}
\end{equation}
where $H_{0}$ is the value of the claim at initial time $t_0=0$ and $H(\mGamma_K, X)$ is the function $H$ applied element-wise to $\mGamma_K$, which, as specified previously, is a column vector of length $N_K$.

Figure \ref{Fig: European_Put} shows the accuracy of put option prices for the GBM and CEV models for a wide range of strikes. The GBM option prices are compared against the Black-Scholes option pricing formula, whereas the CEV prices are compared against the analytical solution originally due to \cite{schroder1989computing} and reformulated in terms of the noncentral chi-squared distribution by \cite{hsu2008constant}. In the graphs, the $x$-axis represents fixed-spot inverse moneyness, which is determined as the variable strike value over the initial asset price, $S_0$.

Even though the Euler scheme is reasonably accurate to start with, the increased accuracy of the Milstein and the simplified weak order 2.0 schemes is evident. For certain strikes the error is reduced by an order of magnitude.

\subsection{Bermudan Option Pricing}

Bermudan option prices are computed using the standard Backward Dynamic Programming Principle (BDPP), an important result from discrete-time optimal stopping theory. \cite{pagesintroduction} reviews the use of the BDPP as applied to grids that result from a quantization.

Once quantization grids and corresponding transition probability matrices have been computed using the RMQ algorithm, the high-level algorithm for Bermudan option pricing may be specified as follows:
\begin{enumerate}
	\item Initialize $\mathbf{h}_K = H(\mGamma_K, X)$
	\item For $k = K - 1,\dots, 1$\\[1mm]
    $\phantom{xxx}$Set $\mathbf{h}_k = \max(H(\mGamma_k, X), e^{-r\Delta t} \mP_{k+1} \mathbf{h}_{k+1})$
	\item Set $H_0=e^{-r\Delta t} \mp_1 \mathbf{h}_{1}$
\end{enumerate}
Here the $\max$ function is applied element-wise with its second argument being the continuation value, which is easily computed as a conditional expectation due to availability of the transition probability matrix at each time step. The initial value of the Bermudan claim is given by $H_0$.

\begin{figure}
     \begin{center}
         \includegraphics[width=\columnwidth]{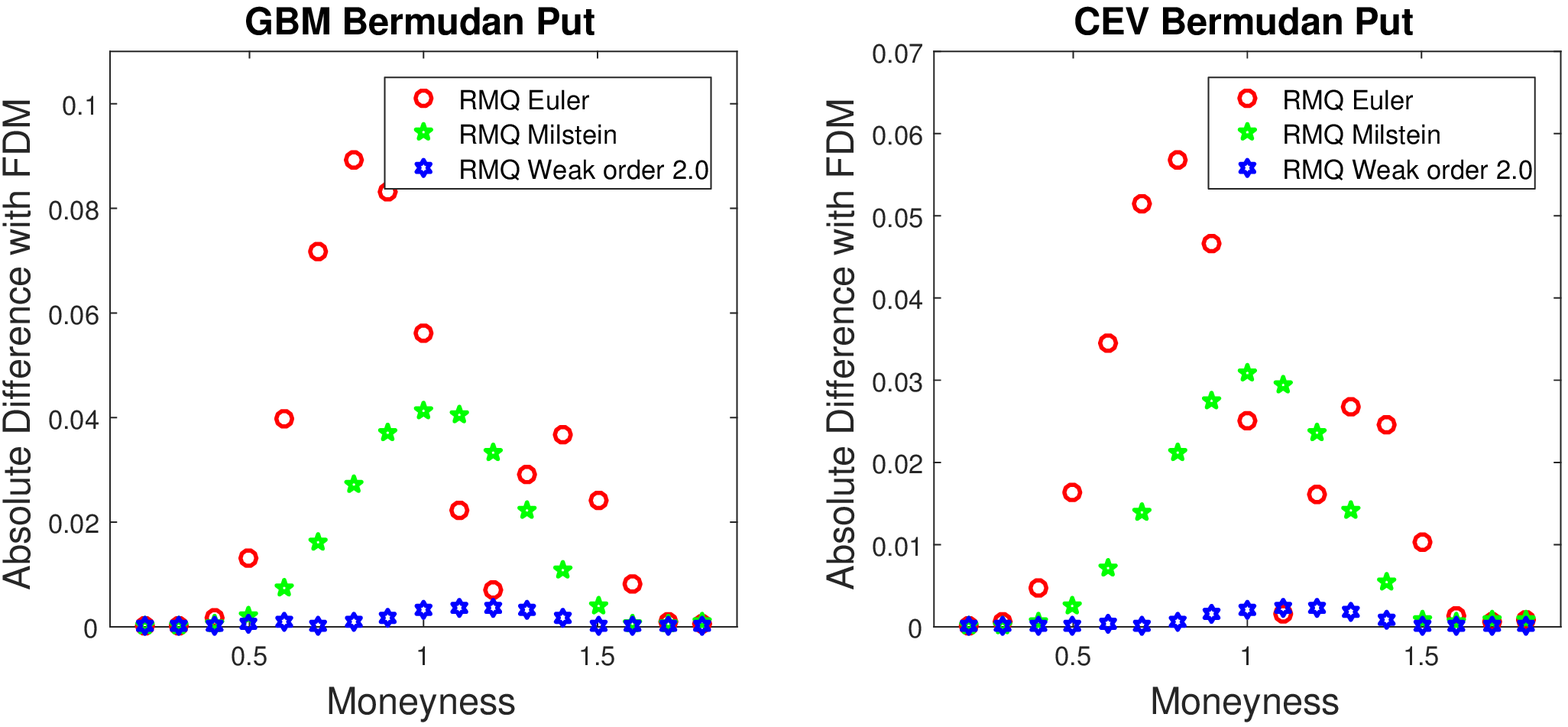}
     \end{center}
     \caption{Accuracy of GBM and CEV Bermudan put option prices, as compared to a high resolution Crank-Nicholson finite difference scheme.}
     \label{Fig: Bermudan}
 \end{figure}

In Figure \ref{Fig: Bermudan} the accuracy of a Bermudan put option with monthly exercise opportunities is shown for the GBM and CEV models. The reference price is computed using a high resolution Crank-Nicholson finite difference scheme using 600 time steps and 800 stock increments, equally spaced between zero and $4\times S_0$.

All three RMQ algorithms result in low absolute errors, with the simplified weak order 2.0 scheme again producing errors that are an order of magnitude smaller.

\subsection{Barrier Option Pricing}

The pricing of barrier options has previously been explored in the context of quantization by \cite{sagna2010pricing}. This work showed that the barrier-crossing approach described in Section 6.4 of \cite{glasserman2003monte} may be applied to marginal quantization using a so-called transition kernel formulation. Using our notation, we now combine this approach with our proposed more accurate schemes and apply it to discretely monitored barrier options.

Consider expression \eqref{Eq: European Options} for pricing European options, which may be re-written as
\[
    H_0\approx e^{-rT}\left(\mp_1\prod_{k=1}^{K-1}\mP_{k+1}\right)H(\mGamma_K,X).
\]
To price a knock-out barrier option the transition probability matrix at each time step in this expression must to be modified to take into account the possibility that the underlying process breaches the barrier. Thus, we rescale the transition probabilities by multiplying them by the probability of not having crossed the barrier.

Let $g(x,y)$ be the probability of transitioning between states $x$ and $y$ without crossing the barrier. If we form an $N_k$ by $N_{k+1}$ matrix of values
\[
    [\mathbf{G}_{k+1}]_{i,j}=g(\cwt{k}{i},\cwt{k+1}{j}),
\]
then $\mP_{k+1}\hprod\mathbf{G}_{k+1}$ defines the transition kernel. The barrier option may then be priced using
\[
    H_0\approx e^{-rT}\left((\mp_1\hprod\mathbf{g}_1)\prod_{k=1}^{K-1} (\mP_{k+1}\hprod\mathbf{G}_{k+1})\right)H(\mGamma_K, X),
\]
where $\mathbf{g}_1=[g(S_0,\cwt{1}{1}),\dots,g(S_0,\cwt{1}{N_1})]$ is a row vector.


In the case of discretely-monitored up-and-out barrier options with barrier level $L$, the function $g$ is given simply as the indicator function
\[
    g(x,y)=\ind{\max(x,y)<L}.
\]
We refer to \cite{glasserman2003monte} and \cite{sagna2010pricing} for the continuous monitoring case using the Rayleigh distribution.



\begin{figure}
     \begin{center}
         \includegraphics[width=\columnwidth]{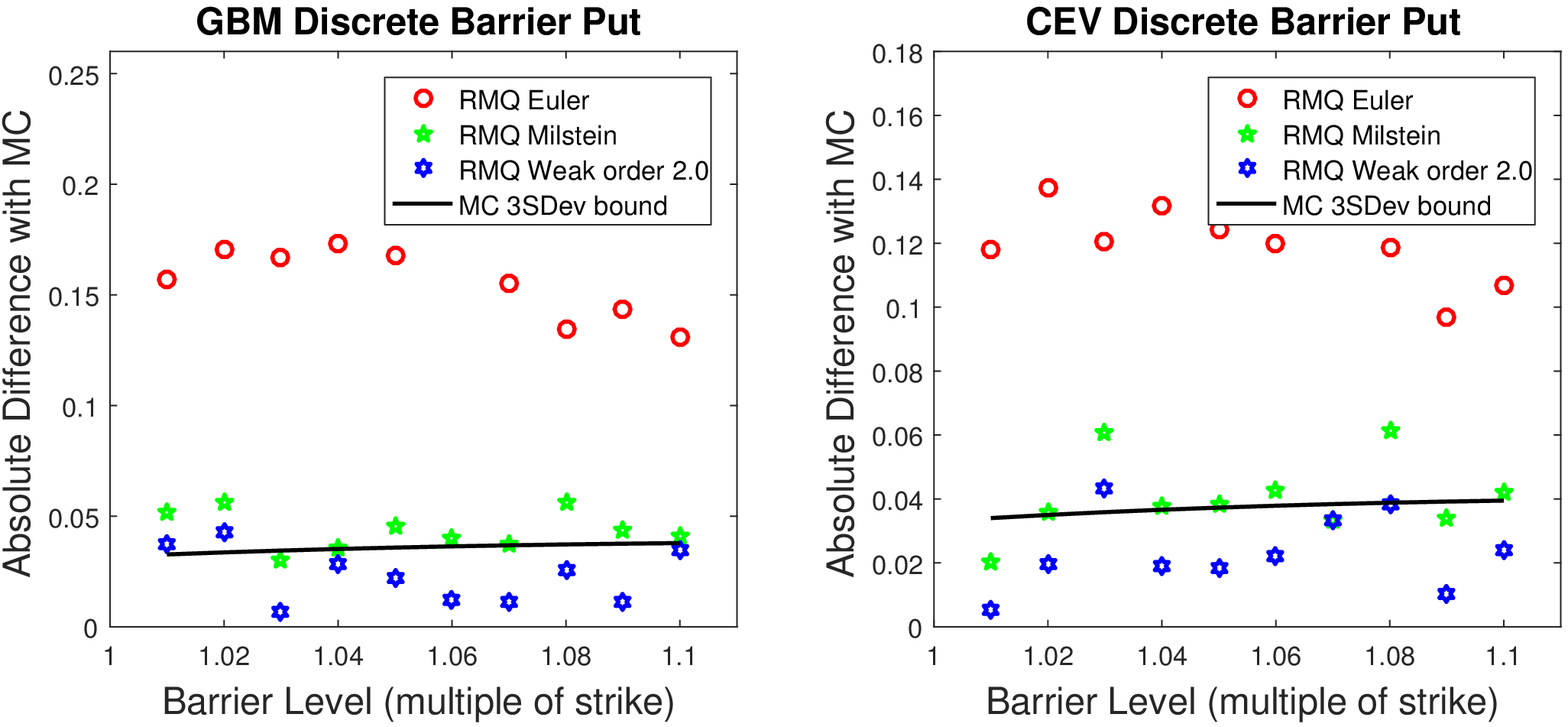}
     \end{center}
     \caption{Accuracy of GBM and CEV discretely monitored up-and-out put option prices, as compared to a Monte Carlo simulation.}
     \label{Fig: Barriers}
 \end{figure}

In Figure \ref{Fig: Barriers} the accuracy of discretely-monitored up-and-out put option prices generated using RMQ is compared to a Monte Carlo implementation under the GBM and CEV models. The barrier levels ($x$-axis) are expressed as multiples of the at-the-money strike. Since we have chosen $K=12$ the barrier is monitored monthly.

The reference prices are provided by a one million path Monte Carlo experiment. The Monte Carlo paths are generated using Euler-Maruyama updates with $1\,200$ time steps, while ensuring that the barriers are only monitored at monthly intervals. We used the exact transition density to generate Monte Carlo samples for GBM to confirm that results were consistent and generating the correct standard deviations.

The results show a similar pattern to those in the previous sections, with an important caveat: the simplified weak order 2.0 scheme produces prices that, for the majority of the barrier values considered, lie within the three standard deviation bound of the million-path Monte Carlo experiment. The other two RMQ schemes are producing results that are statistically significantly incorrect when compared to the Monte Carlo simulation. 

%% file: Sections/Conclusion.tex
In this paper, the recursive marginal quantization methodology of \cite{pages2015recursive} has been extended from the standard Euler-Maruyama scheme to higher-order numerical schemes, specifically the Milstein scheme and the simplified weak order 2.0 scheme of \cite{kloeden1999numerical}. This entailed introducing noncentral chi-squared updates and generalising the formulation of Recursive Marginal Quantization (RMQ).

Is is also shown how to augment the RMQ algorithm in order to implement absorption or reflection at the zero boundary, thus ensuring non-negativity of solutions. This allows RMQ to be applied in important cases, where the algorithm may previously have failed.

Improved approximation of the marginal distributions by the proposed new methods has been demonstrated, using geometric Browninan motion and constant elasticity of variance dynamics. All the schemes were used successfully to price European, Bermudan and discrete barrier options. The pricing results show that once RMQ is implemented with second weak order updates, it provides extremely accurate contingent claim pricing, well suited for the extremely fast calibration of entire derivative books.

%% file: Recursive_Marginal_Quantization.bbl
\begin{thebibliography}{22}
\providecommand{\natexlab}[1]{#1}
\providecommand{\url}[1]{\texttt{#1}}
\expandafter\ifx\csname urlstyle\endcsname\relax
  \providecommand{\doi}[1]{doi: #1}\else
  \providecommand{\doi}{doi: \begingroup \urlstyle{rm}\Url}\fi

\bibitem[Bormetti et~al.(2016)Bormetti, Callegaro, Livieri, and
  Pallavicini]{Bormettietal2016}
G.~Bormetti, G.~Callegaro, G.~Livieri, and A.~Pallavicini.
\newblock A backward {Monte Carlo} approach to exotic option pricing.
\newblock \url{http://ssrn.com/abstract=2686115}, 2016.

\bibitem[Callegaro et~al.(2014)Callegaro, Fiorin, and
  Grasselli]{callegaro2014pricing}
G.~Callegaro, L.~Fiorin, and M.~Grasselli.
\newblock Pricing and calibration in local volatility models via fast
  quantization.
\newblock Available at SSRN 2495829, 2014.

\bibitem[Callegaro et~al.(2015{\natexlab{a}})Callegaro, Fiorin, and
  Grasselli]{Callegaroetal2015a}
G.~Callegaro, L.~Fiorin, and M.~Grasselli.
\newblock Quantized calibration in local volatility.
\newblock \emph{Risk Magazine}, 28:\penalty0 62--67, 2015{\natexlab{a}}.

\bibitem[Callegaro et~al.(2015{\natexlab{b}})Callegaro, Fiorin, and
  Grasselli]{callegaro2015pricing}
G.~Callegaro, L.~Fiorin, and M.~Grasselli.
\newblock Pricing via quantization in stochastic volatility models.
\newblock Available at SSRN 2669734, 2015{\natexlab{b}}.

\bibitem[Du et~al.(1999)Du, Faber, and Gunzburger]{du1999centroidal}
Q.~Du, V.~Faber, and M.~Gunzburger.
\newblock Centroidal {V}oronoi tessellations: {A}pplications and algorithms.
\newblock \emph{SIAM Review}, 41\penalty0 (4):\penalty0 637--676, 1999.

\bibitem[Glasserman(2003)]{glasserman2003monte}
P.~Glasserman.
\newblock \emph{Monte Carlo Methods in Financial Engineering}.
\newblock Springer, 2003.

\bibitem[Graf and Luschgy(2000)]{graf2000foundations}
S.~Graf and H.~Luschgy.
\newblock \emph{Foundations of Quantization for Probability Distributions}.
\newblock Springer, 2000.

\bibitem[Hsu et~al.(2008)Hsu, Lin, and Lee]{hsu2008constant}
Y.-L. Hsu, T.~Lin, and C.~Lee.
\newblock Constant elasticity of variance ({CEV}) option pricing model:
  Integration and detailed derivation.
\newblock \emph{Mathematics and Computers in Simulation}, 79\penalty0
  (1):\penalty0 60--71, 2008.

\bibitem[Kloeden and Platen(1999)]{kloeden1999numerical}
P.~Kloeden and E.~Platen.
\newblock \emph{Numerical Solution of Stochastic Differential Equations}.
\newblock Springer, 1999.

\bibitem[Lindsay and Brecher(2012)]{lindsay2012simulation}
A.~Lindsay and D.~Brecher.
\newblock Simulation of the {CEV} process and the local martingale property.
\newblock \emph{Mathematics and Computers in Simulation}, 82\penalty0
  (5):\penalty0 868--878, 2012.

\bibitem[Lloyd(1982)]{lloyd1982least}
S.~P. Lloyd.
\newblock Least squares quantization in {PCM}.
\newblock \emph{IEEE Transactions on Information Theory}, 28\penalty0
  (2):\penalty0 129--137, 1982.

\bibitem[Lord et~al.(2010)Lord, Koekkoek, and Dijk]{lord2010comparison}
R.~Lord, R.~Koekkoek, and D.~V. Dijk.
\newblock A comparison of biased simulation schemes for stochastic volatility
  models.
\newblock \emph{Quantitative Finance}, 10\penalty0 (2):\penalty0 177--194,
  2010.

\bibitem[Maruyama(1955)]{Maruyama1955}
G.~Maruyama.
\newblock Continuous {M}arkov processes and stochastic equations.
\newblock \emph{Rendiconti del Circolo Matematico di Palermo}, 4\penalty0
  (1):\penalty0 48--90, 1955.

\bibitem[Milstein(1975)]{mil1975approximate}
G.~Milstein.
\newblock Approximate integration of stochastic differential equations.
\newblock \emph{Theory of Probability and Its Applications}, 19\penalty0
  (3):\penalty0 557--562, 1975.

\bibitem[Pag{\`e}s(2014)]{pagesintroduction}
G.~Pag{\`e}s.
\newblock {Introduction to optimal vector quantization and its applications for
  numerics}.
\newblock Technical report, July 2014.
\newblock URL \url{https://hal.archives-ouvertes.fr/hal-01034196}.

\bibitem[{P\`ages} and Pham(2005)]{PagesPham2005}
G.~{P\`ages} and H.~Pham.
\newblock Optimal quantization methods for nonlinear filtering with
  discrete-time observations.
\newblock \emph{Bernoulli}, 11\penalty0 (5):\penalty0 893--932, 2005.

\bibitem[Pag{\`e}s and Sagna(2015)]{pages2015recursive}
G.~Pag{\`e}s and A.~Sagna.
\newblock Recursive marginal quantization of the {E}uler scheme of a diffusion
  process.
\newblock \emph{Applied Mathematical Finance}, 22\penalty0 (5):\penalty0
  463--498, 2015.

\bibitem[{P\`ages} and Wilbertz(2009)]{PagesWilbertz2012}
G.~{P\`ages} and B.~Wilbertz.
\newblock Optimal {Delaunay} and {Voronoi} quantization methods for pricing
  {American} options.
\newblock In R.~Carmona, P.~Hu, P.~Del~Moral, and N.~Oudjane, editors,
  \emph{Numerical methods in Finance}, pages 171--217. Springer, 2009.

\bibitem[{P\`ages} et~al.(2004){P\`ages}, Pham, and
  Printems]{PagesPhamPrintems2004}
G.~{P\`ages}, H.~Pham, and J.~Printems.
\newblock An optimal {M}arkovian quantization algorithm for multidimensional
  stochastic control problems.
\newblock \emph{Stochastics and Dynamics}, 4\penalty0 (4):\penalty0 501--545,
  2004.

\bibitem[Pag{\`e}s et~al.(2004)Pag{\`e}s, Pham, and Printems]{pages2004optimal}
G.~Pag{\`e}s, H.~Pham, and J.~Printems.
\newblock Optimal quantization methods and applications to numerical problems
  in finance.
\newblock In \emph{Handbook of Computational and Numerical Methods in Finance},
  pages 253--297. Springer, 2004.

\bibitem[Sagna(2011)]{sagna2010pricing}
A.~Sagna.
\newblock Pricing of barrier options by marginal functional quantization.
\newblock \emph{Monte Carlo Methods and Applications}, 17\penalty0
  (4):\penalty0 371--398, 2011.

\bibitem[Schroder(1989)]{schroder1989computing}
M.~Schroder.
\newblock Computing the constant elasticity of variance option pricing formula.
\newblock \emph{Journal of Finance}, 44\penalty0 (1):\penalty0 211--219, 1989.

\end{thebibliography}
